\documentclass[epj]{svjour}
\usepackage{graphics}
\usepackage{amsmath,amsfonts,amssymb,amscd}
\usepackage{graphicx}
\usepackage[dvips]{epsfig}
\usepackage{color, colortbl}
\usepackage[table]{xcolor}
\graphicspath{{},{FIGURES/}}
\newcommand{\zh}{\boldsymbol}

\begin{document}
\title{Collapsed heteroclinic snaking near a heteroclinic chain in dragged meniscus problems}
\author{D. Tseluiko$^{\text{a}}$, 
M. Galvagno\thanks{\emph{Note:} The first two authors have equally contributed to the work.},
and U. Thiele
}                     
%
%
\institute{Department of Mathematical Sciences, Loughborough University, Leicestershire, LE11 3TU, UK}
\date{Received: date / Revised version: date}
%
\abstract{ A liquid film is studied that is deposited onto a flat plate
  that is inclined at a constant angle to the horizontal and is
  extracted from a liquid bath at a constant speed. We analyse steady-state
  solutions of a long-wave evolution equation for the film
  thickness. Using centre manifold theory, we first obtain an
  asymptotic expansion of solutions in the bath region. The presence
  of an additional temperature gradient along the plate
  that induces a Marangoni shear stress significantly changes these expansions
  and leads to the presence of logarithmic terms that are absent
  otherwise. Next, we numerically obtain steady solutions and analyse their behaviour as the plate
  velocity is changed. We observe that the bifurcation curve exhibits 
collapsed (or exponential) heteroclinic snaking when the plate inclination angle is above a
  certain critical value. Otherwise, the bifurcation curve is
  monotonic.  The steady profiles along these curves are characterised by a
  foot-like structure that is formed close to the meniscus and is
  preceded by a thin precursor film further up the plate. The length
  of the foot increases along the bifurcation curve.  Finally, we
  prove with a Shilnikov-type method that the snaking behaviour of the bifurcation curves is
  caused by the existence of an infinite number of heteroclinic orbits
  close to a heteroclinic chain that connects in an appropriate
  three-dimensional phase space the fixed point corresponding to the
  precursor film with the fixed point corresponding to the foot and
  then with the fixed point corresponding to the bath.  
} 
\authorrunning{D. Tseluiko, M. Galvagno, and U. Thiele}
\titlerunning{Collapsed heteroclinic snaking in dragged meniscus problems}
\maketitle

\section{Introduction}

Spreading liquids on a surface by pulling a plate out of a liquid bath
is a well known coating process used for industrial applications
\cite{WeRu04}.  In order to gain control over the coating process,
this problem has been studied from an experimental point of view, see,
{\it e.g.},
refs.~\cite{Morey40,Rossum1958,SpSuWi74,SADF07,DFSA08,Maleki2011a},
and also theoretically, see, {\it e.g.},
refs.~\cite{SADF07,LaLe42,Groe70a,Groe70b,Wi81,ZiSnEg09}. Landau and
Levich \cite{LaLe42}, for example, analysed liquid films of constant
thickness coating a vertical plate extracted from a bath of liquid at
low velocities and found that the film thickness scales as $U^{2/3}$,
where $U$ is the velocity of the plate. The asymptotic result of
Landau and Levich was improved by Wilson
\cite{Wi81}. Non-Landau-Levich-type solutions, which satisfy other
scaling laws, were also found, see, for example,
refs.~\cite{SADF07,BCM10,JiAcMu05,Snoe08,ME05}.  In particular,
multiple non-Landau-Levich type solutions were previously
  observed by M\"unch {\it et al.} \cite{ME05} for certain parameter
  values in a similar system, where the role of the plate withdrawal
  is taken by a Marangoni shearing induced by a constant temperature
  gradient on the plate. Related behaviour is also found in coating
problems involving complex fluids. A particular example is the
deposition of line patterns in the process of Langmuir-Blodgett
transfer of a surfactant layer from a bath onto a moving plate
\cite{RiSp1992tsf,KGFC10}. For this system a reduced Cahn-Hilliard type model was employed
to show that the deposition of lines is related to local and global
bifurcations of time-periodic states from a snaking bifurcation curve
of steady-state front solutions \cite{KGFT12}, that in the light of
the present work may be seen as a case of heteroclinic snaking (also
cf.~review~\cite{Thie14} where this is set into the wider context of
deposition patterns).

\rowcolors{1}{lightgray}{}{}
\begin{table*}
\centering
\caption[]{Hierarchy of systems exhibiting collapsed (or
    exponential) snaking behaviour.}
\label{tab:models}
\begin{tabular}{|l|l|c|}
\hline
Authors & Description of scenario&   $\#$ Fixed Points \\ \hline
\rowcolor{white}   Shilnikov \cite{Shil65}   & infinite number of periodic orbits  & 1 fixed point \\ 
\rowcolor{white} Glendinning \& Sparrow \cite{GlSp1984jsp}& approaching a homocline & \\
\hline
\rowcolor{white}   J.~Knobloch $\&$ Wagenknecht \cite{KnoWa05}  & infinite number of homoclines   & 2 fixed points\\ 
\rowcolor{white} Ma, Burke \& E.~Knobloch \cite{MaBK2010pd} & approaching a hetereoclinic cycle & \\
\hline
\rowcolor{white}  Present study   & infinite number of heteroclines  & 3 fixed points\\ 
\rowcolor{white} & approaching a hetereoclinic chain & \\
\hline
\end{tabular}
\end{table*}
\rowcolors{1}{white}{}{}

In the present study, we do not consider Landau-Levich solutions where
the thick drawn film directly connects to the meniscus of the
bath. Instead we focus on a different type of film profiles which show
a foot-like structure of characteristic thickness $h_f$ close to the
meniscus that is preceded by a very thin precursor film of
characteristic thickness $h_p$ further up the plate. They were
recently described for a slip model \cite{SADF07,ZiSnEg09}.  We
  show that for the precursor film model (as known in case of the slip
  model) at inclination angles $\alpha$ below a critical value
  $\alpha_c$, the foot shape is monotonic while for $\alpha>\alpha_c$
  there exist undulations on top of the foot. In both cases we
observe that for each inclination angle foot solutions exist when the
plate velocity is close to a certain limiting velocity, and the closer
the bifurcation curve approaches this limiting value, the larger the
foot length becomes.  The analysis of the bifurcation diagrams of foot
solutions for a suitable solution measure, shows that this classical
physico-chemical problem turns out to be a rich example to illustrate
{collapsed (or exponential)} heteroclinic snaking near a
hetereoclinic chain \cite{noteSnaking}. We demonstrate that the three
regions of the liquid film profile, namely, the precursor film, the
foot and the bath, can be considered as three fixed points $\zh{y}_p$,
$\zh{y}_f$ and $\zh{y}_b$ of an appropriate three-dimensional
dynamical system. The steady film profiles are then described by
heteroclinic orbits connecting points $\zh{y}_p$ and $\zh{y}_b$. Then,
we show that the {collapsed heteroclinic snaking} observed in
the dragged meniscus problem is caused by a perturbation of a
heteroclinic chain that connects $\zh{y}_p$ with $\zh{y}_f$ and
$\zh{y}_f$ with $\zh{y}_b$ that exists for certain parameter values,
provided that fixed points $\zh{y}_p$ and $\zh{y}_b$ have
two-dimensional unstable and two-dimensional stable manifolds,
respectively, and that the fixed point $\zh{y}_f$ is a saddle focus with a
one-dimensional stable manifold and a two-dimensional unstable
manifold.
 
Note that related collapsed snaking behaviour has been
  analysed in  
  systems involving either one fixed point
  \cite{Shil65,Shil67,GlSp1984jsp} or two fixed points \cite{KnoWa05,MaBK2010pd}.
  Table~\ref{tab:models} illustrates that our results form part of a
  hierarchy of such snaking behaviours: Shilnikov (see
refs.~\cite{Shil65,Shil67}) analyses homoclinic orbits to saddle-focus
fixed points in three-di\-men\-sio\-nal dynamical systems that exist for
some value $\beta_0$ of a parameter $\beta$ and demonstrated that if the
fixed point has a one-dimensional unstable manifold and a two
dimensional stable manifold, so that the eigenvalues of the Jacobian
at this point are $\lambda_1$ and $-\lambda_2\pm\mathrm{i}\:\omega,$
where $\lambda_{1,2}$ and $\omega$ are positive real numbers, and if
the saddle index $\delta\equiv\lambda_2/\lambda_1<1$, then in the
neighbourhood of the primary homoclinic orbit there exists an infinite
number of periodic orbits that pass near the fixed point several
times. Moreover, the difference in the periods of these orbit tends
asymptotically to $\pi/\omega$. The perturbation of the structurally
unstable homoclinic orbit leads to a snaking bifurcation diagram
showing the dependence of the period of the orbit versus the
bifurcation parameter $\beta$. This diagram has an infinite but 
countable number of turning points at which the periodic orbits vanish
in saddle-node bifurcations. However, if the saddle index is greater
than unity, then the bifurcation diagram is monotonic. Knobloch
  and Wagenknecht \cite{KnoWa05,KnoWa08} analyse symmetric
  heteroclinic cycles connecting saddle-focus equilibria in reversible
  four-dimensional dynamical systems that arise in a number of
  applications, {\it e.g.}, in models for water waves in horizontal water
  channels \cite{Che00} and in the study of cellular buckling in
  structural mechanics \cite{Hun00}. In these systems the symmetric
  heteroclinic cycle organises the dynamics in an equivalent way to
  the homoclinic solution in Shilnikov's case.  It is found that a
  necessary condition for collapsed snaking in such four-dimensional
  systems is the requirement that one of the involved fixed points is
  a bi-focus \cite{KnoWa05}. Then there exists an infinite number of
  homoclines to the second involved fixed point that all pass a close
  neighbourhood of the bi-focus. The presently studied case is
  equivalent to the cases of Shilnikov and of Knobloch and
  Wagenknecht, however, here a heteroclinic chain between three fixed
  points forms the organising centre of an infinite number of
  heteroclines.
 
The rest of the paper is organised as follows. In sect.~2, we introduce
the model equation. In sect.~3, we analyse asymptotic behaviour of
solutions in the bath region. In sect.~4 we present numerical results
for the steady states and their snaking behaviour in the cases without
and with Marangoni driving.  Section~5 is devoted to an analytical
explanation of the bifurcation diagrams obtained in sect.~4. Finally,
in sect.~6 we present our conclusions.

\section{Model equation}

\begin{figure}[h]
\begin{center}
\includegraphics[width=1.0\hsize]{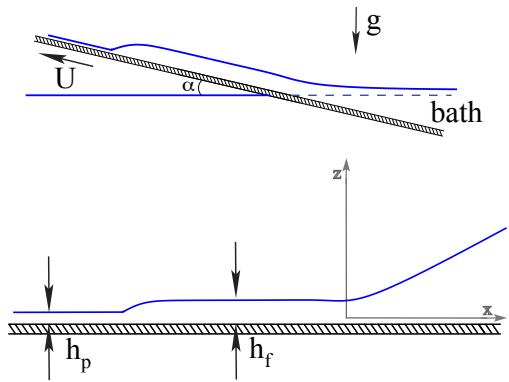}
\caption[System]{Sketch of the problem: Upper panel: An infinitely extended flat plate inclined at an angle $\alpha$ is withdrawn at a constant speed $U$ from a bath of a partially wetting liquid.  Lower panel: Definition of the precursor film height, $h_p$, and the foot film height, $h_f$, for a typical film profile.}
\label{fig:FIG1}
\end{center}
\end{figure}

We consider a flat plate that forms a constant angle with the
horizontal direction and that is being withdrawn from a pool of liquid
at a constant speed. A schematic representation of the system is shown
in fig.~\ref{fig:FIG1}. We introduce a Cartesian coordinate system
$(x,\,z)$ with the $x$-axis pointing downwards along the plate and the
$z$-axis pointing upwards and being perpendicular to the plate. We
assume that the free surface is two-dimensional, with no variations in
the transverse direction. The position of the free surface is given by
the equation $z=h(x,\,t)$, where $t$ denotes time. As a model equation
governing the evolution of the free surface, we use a long-wave
equation derived in refs.~\cite{ODB97,Thie07} from the Navier-Stokes
equations and the corresponding boundary conditions under the
assumptions that the {physical plate inclination angle} is small and the typical longitudinal length scale of free-surface variations is large compared to the typical film thickness:
\begin{eqnarray}
\partial_t h&=&-\partial_x\biggl(\frac{h^3}{3}\partial_x[\partial_x^2h+\Pi(h)]\nonumber\\
&&\qquad\qquad-\frac{h^3}{3}G(\partial_xh-\alpha)-\frac{U}{3}h\biggr). 
\label{eq:thinfilm}
\end{eqnarray}
Here $\alpha$, $U$ and $G$ are the scaled inclination angle of the
plate, the scaled plate velocity and the scaled gravity, respectively,
and the symbols $\partial_t$ and $\partial_x$ denote partial
differentiation with respect to $t$ and $x$, respectively. 
{Note that the scaled angle $\alpha$ as well as the scaled
  equilibrium contact angle are $O(1)$ quantities.}
On the right-hand side, $-\partial_x^2h$ represents the Laplace pressure, $\Pi(h)$ represents the Derjaguin or disjoining pressure (that we will discuss in detail below), 
the term $G\partial_x h$ is due to the hydrostatic-pressure, $-G\alpha$ is due to the $x$-component of gravity and the last term is due to the drag of the plate.

The  interaction between the plate and the non-volatile partially wetting liquid is modelled via the disjoining pressure, which has the dimensional form
\begin{equation}
\widetilde{\Pi}(\tilde h)=\widetilde{\Pi}_1(\tilde h)+\widetilde{\Pi}_2(\tilde h)=-\frac{A}{\tilde h^3}+\frac{B}{\tilde h^6}
\label{eq:derja}
\end{equation} 
consisting of a destabilising long-range van der Waals interaction, $\widetilde{\Pi}_1(\tilde h)=-A/\tilde h^3$ , and a stabilising short-range interaction, $\widetilde{\Pi}_2(\tilde h)=B/\tilde h^6$. Here $\tilde h$ is the dimensional film thickness, and $A$ and $B$ are the Hamaker constants. For $A$ and $B$ positive, on a horizontal plane the disjoining pressure describes partial wetting and characterises a stable precursor film of thickness 
\begin{equation}
h_\mathrm{eq}=(B/A)^{1/3}
\end{equation} 
that may coexist with a meniscus of finite contact angle 
\begin{equation}
\theta_\mathrm{eq}=\sqrt{\frac{3}{5}\frac{A}{\gamma h^2_\mathrm{eq}}},
\end{equation} 
where $\gamma$ is the surface tension coefficient (see refs.~\cite{Thie07,deGe85,StVe09,Thie10} for background information and details).

Equation (\ref{eq:thinfilm}) has been non-dimensionalised using $\ell=\sqrt{3/5}\:h_\mathrm{eq}/\theta_\mathrm{eq}$ as the length scale in the $x$-direction, $h_\mathrm{eq}$ as the length scale in the $z$-direction and $\tau\!\!=\!\!(9\eta h_\mathrm{eq})/(25\gamma\theta^{4}_\mathrm{eq})$ as the time scale, where $\eta$ is the viscosity of the liquid. Note that with this non-dimensionalisation the dimensionless disjoining pressure has the form
\begin{equation}
{\Pi}( h)={\Pi}_1( h)+{\Pi}_2( h)=-\frac{1}{ h^3}+\frac{1}{ h^6}.
\label{eq:derja_nondim}
\end{equation} 
The scaled velocity, gravity number and the inclination angle are given by
\begin{equation}
U=\frac{3\tau}{\ell}u,\quad G=\frac{\rho g h_\mathrm{eq}^4}{A},\quad \alpha=\frac{\ell }{h_\mathrm{eq}}\tilde \alpha,
\end{equation}
respectively, where $\rho$ is the density of the liquid and $g$ is gravity and $u$ and $\tilde \alpha$ are the dimensional plate velocity and the plate inclination angle, respectively.

Note that additional physical effects can be included into the model
presented above. One extension that is interesting for reasons that will become clear later, is the inclusion of a term quadratic in $h$ in the flux on the right-hand side of eq.~(\ref{eq:thinfilm}). This can be obtained, for example, by assuming that there is an additional constant temperature gradient along the plate, see refs.~\cite{Cazabat1990,ME05,ScNiSt10,ScNiSt12} for more details. Inclusion of this effect into the present model results in 
\begin{eqnarray}
\partial_t
h&=&-\partial_x\biggl(\frac{h^3}{3}\partial_x[\partial_x^2h+\Pi(h)]\nonumber\\
&&\qquad-\frac{h^3}{3}G(\partial_xh-\alpha)
-\frac{\Omega}{3}h^2 -\frac{U}{3}h\biggr), 
\label{eq:thinfilm_temperature}
\end{eqnarray}
where $\Omega$ is a dimensionless number representing the temperature gradient along the plate. 

Finally, we discuss boundary conditions. First, we assume that $h$
tends to an undetermined constant value ({\it e.g.}, at equilibrium
the precursor film thickness) as $x\rightarrow-\infty$ and its
derivatives tend to zero as $x\rightarrow-\infty$. Second, we assume
that $h_x=\alpha+o(1)$ as $x\rightarrow\infty$, which means that the
slope of the free surface of the bath approaches the horizontal direction far away from the plate. The asymptotic behaviour of $h$ as $x\rightarrow\infty$ will be analysed in more detail in the next section.

\section{Asymptotic behaviour of solutions at infinity}

In what follows, we will analyse steady-state solutions of eq.~(\ref{eq:thinfilm_temperature}), {\it i.e.}, solutions that  satisfy the equation
\begin{equation}
h^3[h''+\Pi(h)]'-Gh^3(h'-\alpha)-\Omega h^2-Uh+C_0=0,
\label{eq:steady_eq}
\end{equation}
where now $h$ is a function of $x$ only and primes denote differentiation with respect to $x$. Here, $C_0$ is a constant of integration and represents the flux. Note that $C_0$ is in fact not an independent parameter but is determined as part of the solution of the boundary-value problem consisting of eq.~(\ref{eq:steady_eq}) and four boundary conditions that will be discussed in the next section.

Following a proposal of ref.~\cite{ME05}, we introduce variables $y_1=1/h$, $y_2=h'$ and $y_3=h''$, and convert the steady-state equation (\ref{eq:steady_eq}) into a three-dimensional dynamical system:
\begin{eqnarray}
y_1'&=&-y_1^2y_2,\label{eq:y1_prime}\\
y_2'&=&y_3,\\
y_3'&=&(6y_1^7-3y_1^4)y_2+Gy_2+Uy_1^2\nonumber\\
&&+\Omega y_1-C_0y_1^3-G\alpha.\label{eq:y3_prime}
\end{eqnarray}
Note that the transformation $y_1=1/h$ is used to obtain a new fixed
point corresponding to the bath, namely the point
$\zh{y}_b=(0,\,\alpha,\,0)$, beside other fixed points, two of which,
$\zh{y}_f=(1/h_f,\,0,\,0)$ and $\zh{y}_p=(1/h_p,\,0,\,0)$, correspond
to the foot and the precursor film, respectively. {For  a more detailed analysis of the fixed points, see the beginning of sect.~5.} 

To analyse the stability of the fixed point $\zh{y_b}$, we first compute the Jacobian at this point:
\begin{equation}
\zh{J}_{\zh{y}_b}=\left(
\begin{array}{ccc}
0 & 0 & 0\\
0 & 0 & 1\\
\Omega & G & 0
\end{array}
\right).
\end{equation}
The eigenvalues are $0$, $\pm G^{1/2}$ and the corresponding eigenvectors are $(G,\,-\Omega,\,0)$, $(0,\,\pm G^{-1/2},\,1)$. So there is a one-dimensional centre (or critical) eigenspace, a  one-di\-me\-n\-si\-o\-nal stable eigenspace  and a one-dimensional unstable eigenspace given by
\begin{eqnarray}
T^c_{\zh{y}_b}&=&\mathrm{span}\{(G,\,-\Omega,\,0)\},\\
T^s_{\zh{y}_b}&=&\mathrm{span}\{(0,\,-G^{-1/2},\,1)\},\\
T^u_{\zh{y}_b}&=&\mathrm{span}\{(0,\,G^{-1/2},\,1)\},
\end{eqnarray}
respectively.

To determine the asymptotic behaviour of $h$ as $x\rightarrow\infty$,
we analyse the centre manifold of $\zh{y}_b$, which we denote by
$W^c_{\zh{y}_b}$. This is an invariant manifold whose tangent space at
$\zh{y}_b$ is $T^c_{\zh{y}_b}$. The existence of a centre manifold is
provided by the centre manifold theorem (see, {\it e.g.}, theorem~1,
p.~4 in ref.~\cite{Carr_1981}, theorem~5.1, p.~152 in
ref.~\cite{Kuz98}). For simplicity, we use the substitution $z_1=y_1$,
$z_2=y_2-\alpha$, $z_3=y_3$. 
In vector notation, the dynamical system takes the form
\begin{equation}
\zh{z}'=\zh{f}(\zh{z}),
\label{eq:z'}
\end{equation}
where 
$\zh{f}(\zh{z})=(f_1(\zh{z}),\,f_2(\zh{z}),\,f_3(\zh{z}))^T$ and
\begin{eqnarray}
\!\!\!\!\!\!f_1(\zh{z})=f_1(z_1,\,z_2,\,z_3)&=&-z_1^2(z_2+\alpha),\\
\!\!\!\!\!\!f_2(\zh{z})=f_2(z_1,\,z_2,\,z_3)&=&z_3,\\
\!\!\!\!\!\!f_3(\zh{z})=f_3(z_1,\,z_2,\,z_3)&=&(6z_1^7-3z_1^4)(z_2+\alpha)+Gz_2\nonumber\\
&&+Uz_1^2+\Omega z_1-C_0z_1^3.\label{eq:f3}
\end{eqnarray}
The fixed point corresponding to the bath is then $\zh{z}_b=(0,\,0,\,0)$. Next, we rewrite the system of ordinary differential equations (\ref{eq:z'}) in its eigenbasis at $\zh{z}_b$, {\it i.e.}, we use the change of variables $\zh{u}=\zh{B}^{-1}\zh{z}$, where $\zh{B}$ is the matrix having the eigenvectors of the Jacobian as its columns,
\begin{equation}
\zh{B}=\left(
\begin{array}{ccc}
G & 0 & 0\\
-\Omega & G^{-1/2} & -G^{-1/2}\\
0 & 1 & 1
\end{array}
\right),
\end{equation}
and obtain the system 
\begin{eqnarray}
\zh{u}'=\zh{g}(\zh{u})\equiv \zh{B}^{-1}\zh{f}(\zh{B}\zh{u}),
\end{eqnarray}
which can be written in the form 
\begin{eqnarray}
&&\xi'=\psi(\xi,\zh{\eta}),\label{eq:xi}\\
&&\zh{\eta}'=\zh{C}\zh{\eta}+\zh{\varphi}(\xi,\zh{\eta})\label{eq:eta},
\end{eqnarray}
where $\xi$ denotes the first component of $\zh{u}$ and $\zh{\eta}=(\eta_1,\,\eta_2)^T$ consist of the second and the third components of $\zh{u}$ ({\it i.e.}, $\xi\equiv u_1$, $\eta_1\equiv u_2$ and $\eta_2\equiv u_3$), $\psi$ and $\zh{\varphi}$ have Taylor expansions that start  with quadratic or even higher order terms and $\zh{C}$ is the matrix
\begin{equation}
\zh{C}=\left(\begin{array}{cc}
G^{1/2} & 0\\
0 & -G^{1/2}
\end{array}
\right).
\end{equation}
After some algebra, we find
\begin{eqnarray}
\psi(\xi,\zh{\eta})&=&G{\Omega}\xi ^3-G{\alpha}\xi ^2-G^{1/2}\xi ^2\eta_1 +G^{1/2}\xi ^2\eta_2,\\
\varphi_1(\xi,\zh{\eta})&=&-3\,{G}^{7}\Omega \,{\xi}^{8}+3\,{G}^{7}\alpha \,{\xi}^{7}+3\,{G}^{13/2}{\xi}^{7}\eta_1\nonumber\\
&&-3\,{G}^{13/2}{\xi}^{7}\eta_2+\frac{3}{2}\,{G}^{4}\Omega \,{\xi}^{5}-\frac{3}{2}\,{G}^{4}\alpha \,{\xi}^{4}\nonumber\\
&&-\frac{3}{2}\,{G}^{7/2}{\xi}^{4}\eta_1+\frac{3}{2}\,{G}^{7/2}{\xi}^{4}\eta_2-\frac{1}{2}\,C_0\,{G}^{3}{\xi}^{3}\nonumber\\
&&+\frac{1}{2}\,{G}^{3/2}\Omega ^{2}{\xi}^{3}-\frac{1}{2}\,{G}^{3/2}\Omega \,\alpha \,{\xi}^{2}+\frac{1}{2}\,U{G}^{2}{\xi}^{2}\nonumber\\
&&-\frac{1}{2}\,G\Omega \,{\xi}^{2}\eta_1+\frac{1}{2}\,G\Omega \,{\xi}^{2}\eta_2,\\
\varphi_2(\xi,\zh{\eta})&=&- {G}^{1/2}\eta_2-3\,{G}^{7}\Omega \,{\xi}^{8}+3\,{G}^{7}\alpha \,{\xi}^{7}\nonumber\\
&&+3\,{G}^{13/2}{\xi}^{7}\eta_1-3\,{G}^{13/2}{\xi}^{7}\eta_2+\frac{3}{2}\,{G}^{4}\Omega \,{\xi}^{5}\nonumber\\
&&-\frac{3}{2}\,{G}^{4}\alpha \,{\xi}^{4}-\frac{3}{2}\,{G}^{7/2}{\xi}^{4}\eta_1+\frac{3}{2}\,{G}^{7/2}{\xi}^{4}\eta_2\nonumber\\
&&-\frac{1}{2}\,{C_0}\,{G}^{3}{\xi}^{3}-\frac{1}{2}\,{G}^{3/2}\Omega ^{2}{\xi}^{3}+\frac{1}{2}\,{G}^{3/2}\Omega \,\alpha \,{\xi}^{2}\nonumber\\
&&+\frac{1}{2}\,U{G}^{2}{\xi}^{2}+\frac{1}{2}\,G\Omega \,{\xi}^{2}\eta_1-\frac{1}{2}\,G\Omega \,{\xi}^{2}\eta_2.\label{eq:g3}
\end{eqnarray}

Near the origin, $\zh{z_b}$, when $|\xi|<\delta$ for some positive $\delta$, the centre manifold in the $(\xi,\eta_1,\eta_2)$-space can be represented by the equations 
$\eta_1=g_1(\xi)$, $\eta_2=g_2(\xi)$, where $g_1$ and $g_2$ are in $C^2$. Moreover, near the origin system (\ref{eq:xi}), (\ref{eq:eta}) is topologically equivalent to the system
\begin{eqnarray}
&&\xi'=\psi(\xi,\zh{g}(\xi)),\label{eq:xi_center}\\
&&\zh{\eta}'=\zh{C}\zh{\eta}\label{eq:eta_center}.
\end{eqnarray}
where the first equation represents the restriction of the flow to its centre manifold (see, {\it e.g.}, theorem~1, p.~4 in ref.~\cite{Carr_1981}, theorem~5.2, p. 155 in ref.~\cite{Kuz98}). 

The centre manifold can be approximated to any degree of accuracy. According to theorem~3, p.~5 in ref.~\cite{Carr_1981}, `test' functions $\phi_1$ and $\phi_2$ approximate the centre manifold with accuracy $O(|\xi|^q)
$, namely,
\begin{equation}
|g_1(\xi)-\phi_1(\xi)|=O(|\xi|^q),\quad |g_2(\xi)-\phi_2(\xi)|=O(|\xi|^q)
\end{equation}
as $\xi\rightarrow 0$, provided that $\phi_i(0)=0$, $\phi_i'(0)=0$, $i=1,\,2$ and $\zh{M}[\zh{\phi}](\xi)=O(|\xi|^q)$ as $\xi\rightarrow 0$, where $\zh{M}$ is the operator defined by
\begin{equation}
\zh{M}[\zh{\phi}](\xi)=\zh{\phi}'(\xi)\psi(\xi,\,\zh{\phi}(\xi))-\zh{C}\zh{\phi}(\xi)-\zh{\varphi}(\xi,\,\zh{\phi}(\xi)).
\end{equation}
The centre manifold can now be obtained by seeking for $\phi_1(\xi)$ and $\phi_2(\xi)$ in the form of polynomials in $\xi$ and requiring that the coefficients of the expansion of $\zh{M}[\zh{\phi}](\xi)$ in Taylor series vanish at zeroth order, first order, second order, etc. Using this procedure, we can find the Taylor series expansions of $g_1$ and $g_2$:
\begin{eqnarray}
g_1(\xi)&\!\!\!=\!\!\!&\left(\frac{1}{2}\,G\Omega \,\alpha -\frac{1}{2}\,{G}^{3/2}U\right)\xi^2\nonumber\\
&&+\left({G}^{2}U\alpha -{G}^{3/2}\Omega \,{\alpha }^{2}-\frac{1}{2}\,G\Omega ^{2}+\frac{1}{2}\,{G}^{5/2} C_0 \right)\xi^3\nonumber\\
&&-\biggl(\frac{3}{2}\,{G}^{3}\alpha \, C_0 -3\,{G}^{2}{\alpha }^{3}\Omega +\frac{3}{2}\,{G}^{2}\Omega \,U-3\,{G}^{5/2}{\alpha }^{2}U\nonumber\\
&&\qquad-\frac{3}{2}\,{G}^{7/2}\alpha -\frac{5}{2}\,{G}^{3/2}\alpha \,\Omega ^{2}\biggr)\xi^4+\cdots,\label{eq:g1}\\
g_2(\xi)&\!\!\!=\!\!\!&\left(\frac{1}{2}\,G\Omega \,\alpha +\frac{1}{2}\,{G}^{3/2}U\right)\xi^2\nonumber\\
&&+\left({G}^{2}U\alpha +{G}^{3/2}\Omega \,{\alpha }^{2}-\frac{1}{2}\,G\Omega ^{2}-\frac{1}{2}\,{G}^{5/2} C_0 \right)\xi^3\nonumber\\
&&-\biggl(\frac{3}{2}\,{G}^{3}\alpha \, C_0 -3\,{G}^{2}{\alpha }^{3}\Omega -3\,{G}^{5/2}{\alpha }^{2}U+\frac{3}{2}\,{G}^{2}\Omega \,U\nonumber\\
&&\qquad+\frac{3}{2}\,{G}^{7/2}\alpha +\frac{5}{2}\,{G}^{3/2}\alpha \,\Omega^{2}\biggr)\xi^4+\cdots.\label{eq:g2}
\end{eqnarray}
Let $g_i^{(k)}(\xi)$, $i=1,\,2$, be the Taylor polynomial for $g_i(\xi)$ of degree $k$. Then $g_i(\xi)=g_i^{(k)}(\xi)+O(|\xi|^{k+1})$,  $i=1,\,2$, and $\zh{M}[\zh{g}^{(k)}](\xi)=O(|\xi|^{k+1})$ as $\xi\rightarrow 0$. The dynamics on the centre manifold is therefore governed by the equation
\begin{eqnarray}
\xi'&=&\psi(\xi,\zh{g}^{(k)}(\xi))+O(|\xi|^{k+3})\nonumber\\
&=&G{\Omega}\xi ^3-G{\alpha}\xi ^2-G^{1/2}\xi ^2g_1^{(k)}(\xi)\nonumber\\
&& +G^{1/2}\xi ^2g_2^{(k)}(\xi)+O(|\xi|^{k+3}).\label{eq:centre_dynamics}
\end{eqnarray}
Substituting eq.~(\ref{eq:g1}) and eq.~(\ref{eq:g2}) into eq.~(\ref{eq:centre_dynamics}), we find
\begin{eqnarray}
\xi'&=&-G\alpha\xi^2+G\Omega\xi^3+UG^2\xi^4-(C_0 G^3-2G^2\Omega\alpha^2)\xi^5\nonumber\\
&&\qquad+(6 G^3 U\alpha^2-3G^4\alpha-5G^2\Omega^2\alpha)\xi^6+\cdots.
\end{eqnarray}
Taking into account the fact that $\xi=z_1/G$, we obtain
\begin{eqnarray}
z_1'&&=-\alpha z_1^2+\frac{\Omega}{G}z_1^3+\frac{U}{G}z_1^4-\biggl(\frac{C_0}{G}-\frac{2\Omega\alpha^2}{G^2}\biggr)z_1^5\nonumber\\
&&+\biggl( \frac{6U\alpha^2}{G^2}-\frac{3\alpha}{G}-\frac{5\Omega^2\alpha}{G^3}\biggr)z_1^6+\cdots.
\end{eqnarray}
Rewriting this in terms of $h$, we get
\begin{eqnarray}
h'&&=\alpha-\frac{\Omega}{G}h^{-1}-\frac{U}{G}h^{-2}+\biggl(\frac{C_0}{G}-\frac{2\Omega\alpha^2}{G^2}\biggr)h^{-3}\nonumber\\
&&\quad-\biggl( \frac{6U\alpha^2}{G^2}-\frac{3\alpha}{G}-\frac{5\Omega^2\alpha}{G^3}\biggr)h^{-4}+\cdots\label{eq:h_ode}
\end{eqnarray}
as $h\rightarrow\infty$.

We seek for a solution for $h$ whose slope approaches that of the line corresponding to the horizontal direction as $x\rightarrow\infty$. In the chosen system of coordinates, the line corresponding to the horizontal direction has the slope $\alpha$. So we seek for a solution satisfying $h'(x)=\alpha+o(1)$ as $x\rightarrow\infty$. This can also be written in the form 
\begin{equation}\label{eq:h_1}
h(x)=\alpha x+o(x)\quad \text{as}\quad x\rightarrow\infty.
\end{equation} 

Substituting eq.~(\ref{eq:h_1}) into eq.~(\ref{eq:h_ode}), we obtain
\begin{equation}
h'=\alpha-\frac{\Omega}{\alpha G}x^{-1}+o(x^{-1}),
\end{equation}
which implies
\begin{equation}\label{eq:h_2}
h=\alpha x-\frac{\Omega}{\alpha G}\log x+o(\log x).
\end{equation}
Substituting eq.~(\ref{eq:h_2}) into eq.~(\ref{eq:h_ode}), we find
\begin{equation}
h'=\alpha-\frac{\Omega}{\alpha G}x^{-1}-\frac{\Omega^2}{\alpha^3 G^2}x^{-2}\log x+o(x^{-2}\log x),
\end{equation}
which implies
\begin{equation}\label{eq:h_3}
h=\alpha x-\frac{\Omega}{\alpha G}\log x+\frac{\Omega^2}{\alpha^3 G^2}x^{-1}\log x+o(x^{-1}\log x).
\end{equation}
In principle, any constant of integration can be added to this expression, and this reflects the fact that there is translational invariance in the problem, {\it i.e.}, if $h(x)$ is a solution of eq.~(\ref{eq:steady_eq}), then a profile obtained by shifting $h(x)$ along the $x$-axis is also a solution of this equation. Without loss of generality, we choose the constant of integration to be zero, which breaks this translational invariance and allows selecting a unique solution from the infinite set of solutions.

\begin{figure*}

\includegraphics[width=0.49\hsize]{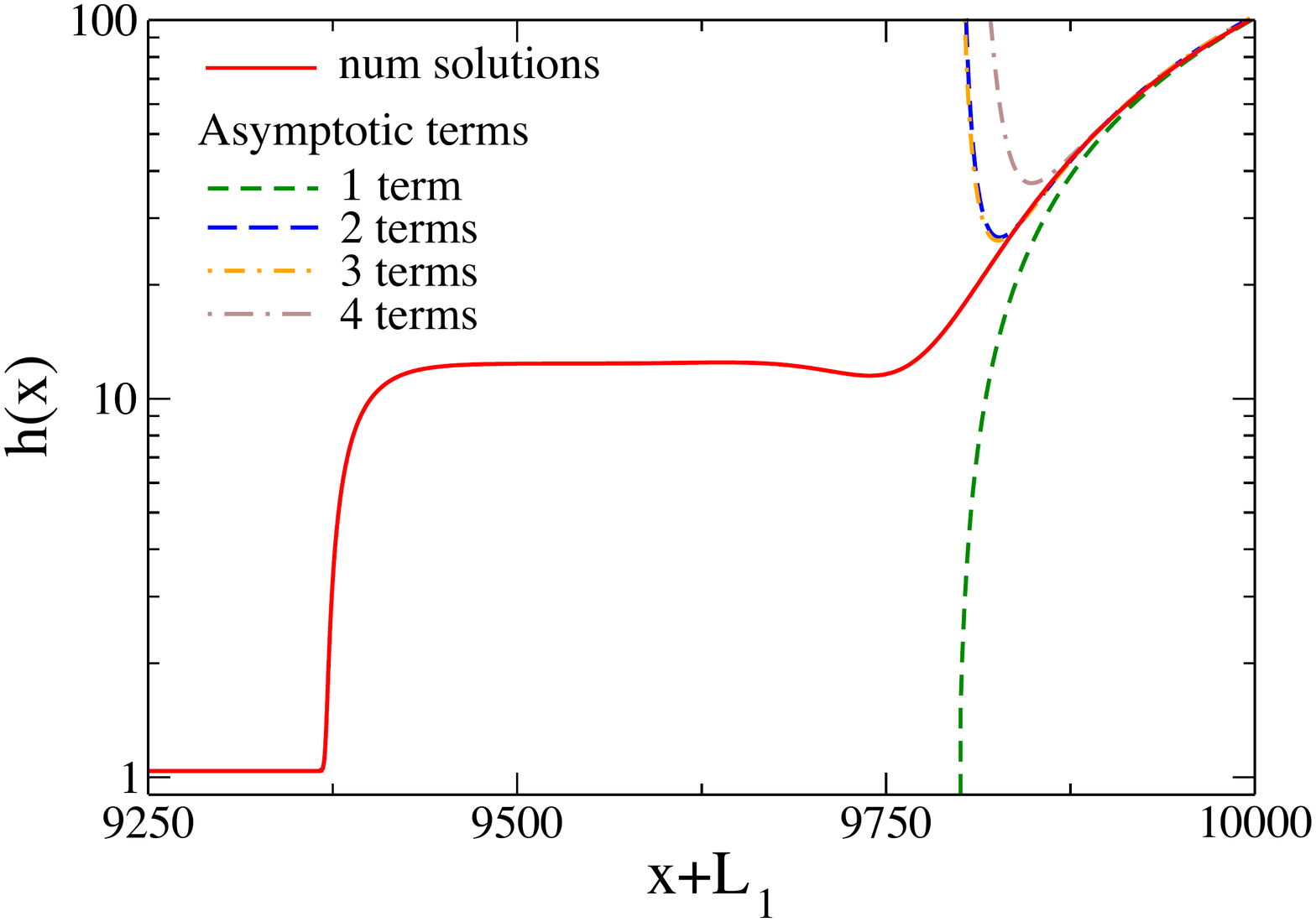}
\includegraphics[width=0.49\hsize]{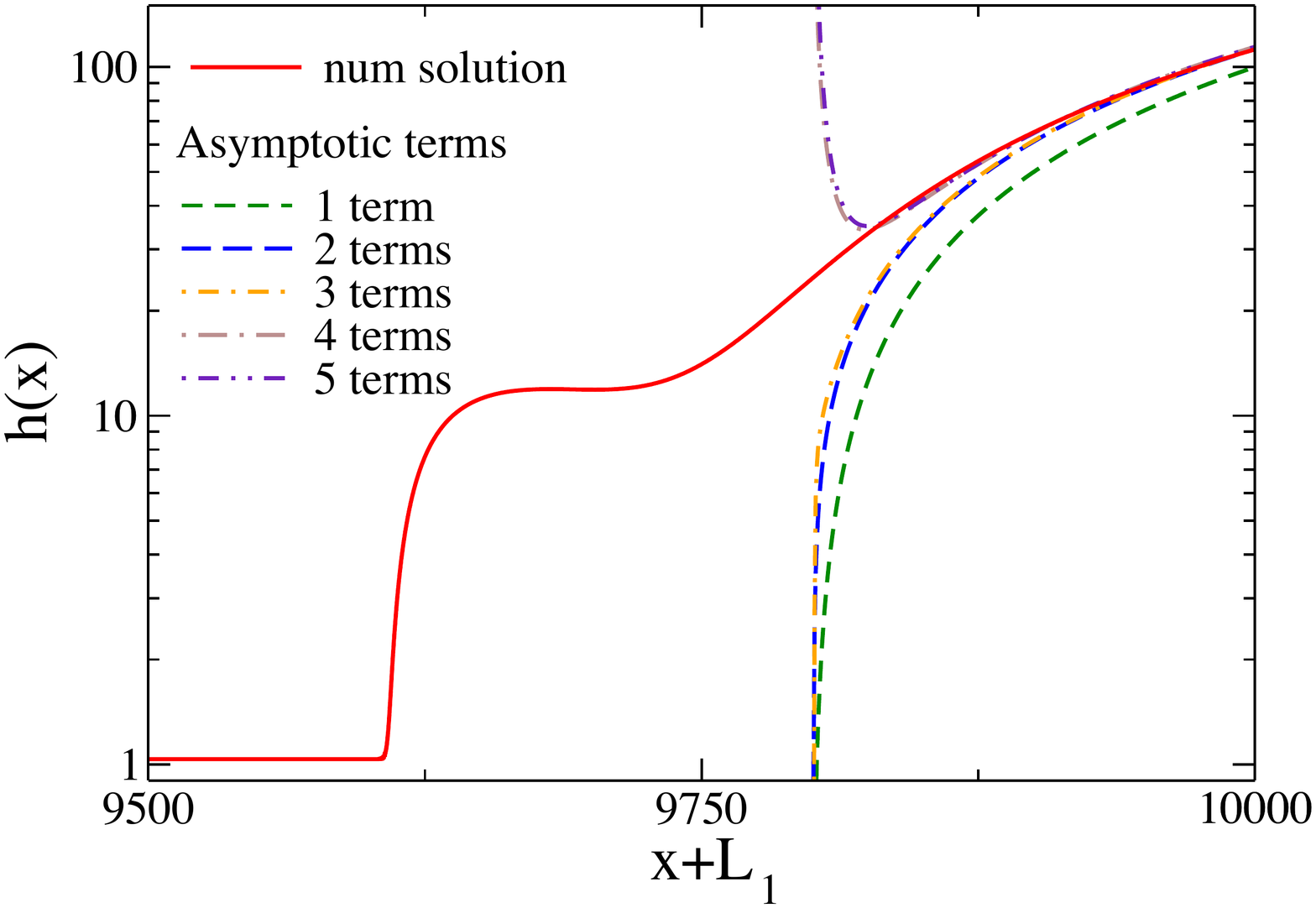}
\caption[Asymptotics]{Left panel: Comparison between a numerical solution for $\Omega=0$ when $\alpha=0.5$ and $U=0.084$ and the expansion for $h(x)$ given by eq.~(\ref{eq:ansatz}) with 1-4 terms. 
Right panel: Comparison between a numerical solution for $\Omega=0.001$ when $\alpha=0.5$ at $U=0.076$ and the expansion for $h(x)$ given by eq.~(\ref{eq:h_4}) with 1-5 terms.  $L_1=9800$, $L_2=200$.}
\label{fig:FIG3}
\end{figure*}

Substituting eq.~(\ref{eq:h_3}) into eq.~(\ref{eq:h_ode}), we find
\begin{eqnarray}
h'&=&\alpha-\frac{\Omega}{\alpha G}x^{-1}-\frac{\Omega^2}{\alpha^3 G^2}x^{-2}\log x\nonumber\\
&&-\frac{U}{\alpha^2G}x^{-2}-\frac{\Omega^3}{\alpha^5 G^3}x^{-3}\log^2 x\nonumber\\
&&+\frac{\Omega^3}{\alpha^5 G^3}x^{-3}\log x+o(x^{-3}\log x),
\end{eqnarray}
which implies
\begin{eqnarray}\label{eq:h_4}
h&=&\alpha x-\frac{\Omega}{\alpha G}\log x+\frac{\Omega^2}{\alpha^3 G^2}x^{-1}\log x\\
&&+\biggl(\frac{\Omega^2}{\alpha^3 G^2}+\frac{U}{\alpha^2 G}\biggr)x^{-1} \nonumber\\
&&-\frac{\Omega^3}{2\alpha^5 G^3}x^{-2}\log^2 x  +o(x^{-2}\log x).
\end{eqnarray}
The procedure described above can be continued to obtain more terms in
the asymptotic expansion of $h$ as $x\rightarrow\infty$. Note that all
the terms in this expansion, except the first two, will be of the form
$x^{-m}\log^n x$, where $m$ is a positive integer and $n$ is a
non-negative integer. It should also be noted that the presence of the
logarithmic terms in the asymptotic expansion of $h$ is wholly due to
the quadratic contribution to the flux in
eq.~(\ref{eq:thinfilm_temperature}) that here
results from a lateral temperature gradient. Without this term, {\it i.e.}, for $\Omega=0$, the expansion (\ref{eq:h_ode}) for $h'$ does not contain the term proportional to $h^{-1}$. This implies that after substituting $h(x)=\alpha x+o(x)$ in this expansion, no term proportional to $x^{-1}$ will appear, and, therefore, integration will not lead to the appearance of a logarithmic term. In fact, it is straightforward to see that for $\Omega=0$ an appropriate ansatz for $h$ is
\begin{equation}\label{eq:ansatz}
h\sim \alpha x+D_1 x^{-1}+D_2 x^{-2}+D_3 x^{-3}+\cdots,
\end{equation}
implying that
\begin{eqnarray}
D_1&=&\frac{U}{\alpha^2 G}, \qquad D_2=-\frac{C_0}{2\alpha^3G},\nonumber\\ 
D_3&=&-\frac{1}{3}\biggl(\frac{2U^2}{\alpha^5G}+\frac{3}{\alpha^3G}-\frac{6U}{\alpha^2G^2}\biggr),\,\, \ldots\,.
\label{eq:asympt_coeff}
\end{eqnarray}

Note that the presence of a logarithmic term in the asymptotic behaviour of $h$ was also observed by M\"unch and Evans~\cite{ME05} in a related problem of a liquid film driven out of a meniscus by a thermally induced Marangoni shear stress  onto a nearly horizontal fixed plane. They find the following asymptotic behaviour of the solution, given with our definition of the coordinate system:
\begin{equation}
h(x)\sim h_0(x) +D_0+D_1 \exp(-D^{1/2}x)\quad\text{as}\quad x\rightarrow\infty,
\end{equation}
where $h_0=x/D-\log x+o(1)$, $D$ is the parameter measuring the relative importance of the normal component of gravity and $D_0$ and $D_1$ are arbitrary constants. 
The constant $D_0$ reflects the fact that there is translational invariance in the problem and it can be set to zero without loss of generality. An analysis performed along the lines indicated above shows that a more complete expansion has the form
\begin{eqnarray}
h(x)\sim \frac{x}{D}-\log x+D\:x^{-1}\log x+D{x^{-1}}\nonumber\\
+\frac{D^2}{2}x^{-2}\log^2 x+\cdots.
\end{eqnarray}
Note that there is no need to include the exponentially small term as
it is asymptotically smaller than all the other terms of the expansion.

\begin{figure}
\includegraphics[width=1\hsize]{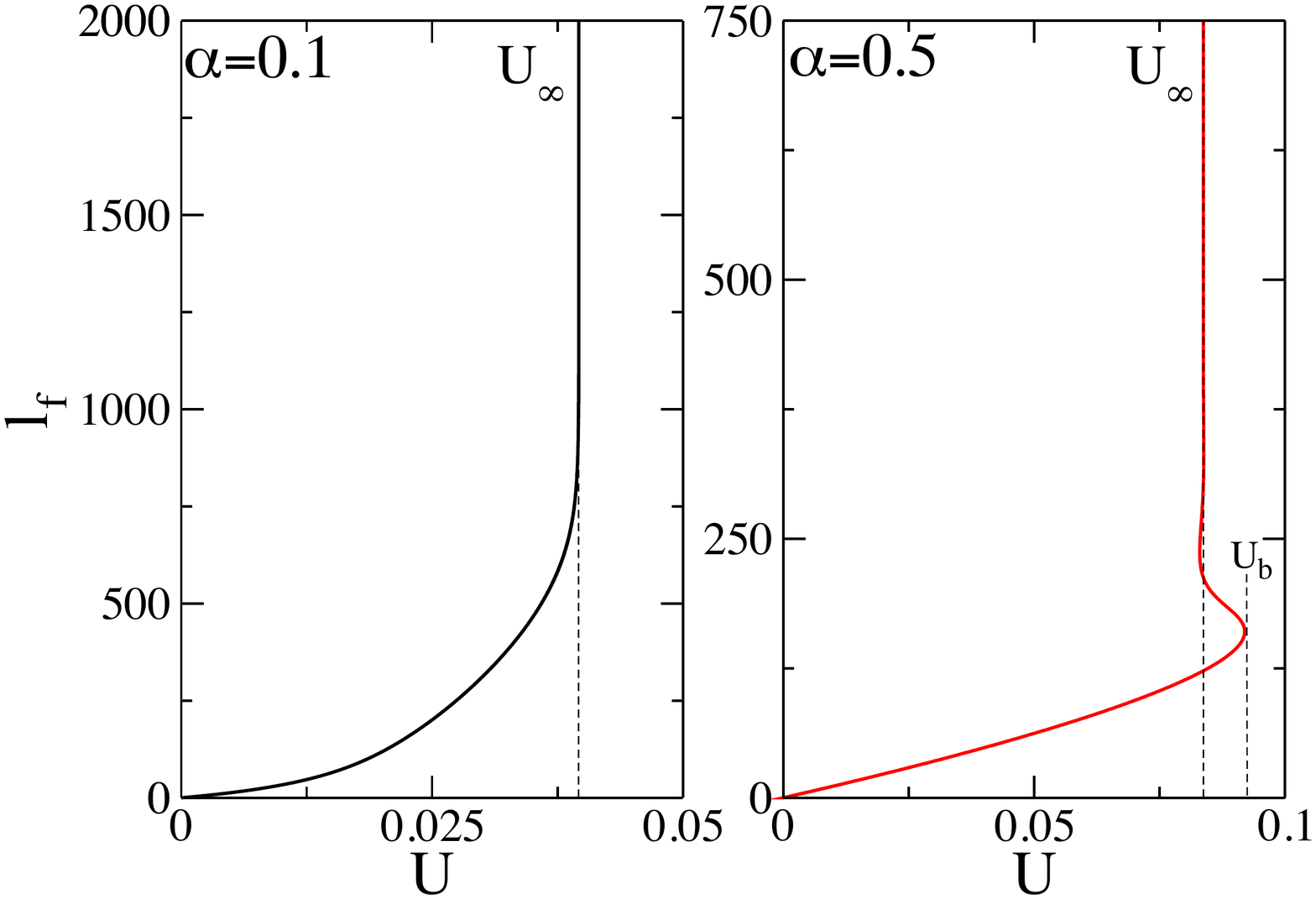}\\[-1cm]
\includegraphics[width=1\hsize]{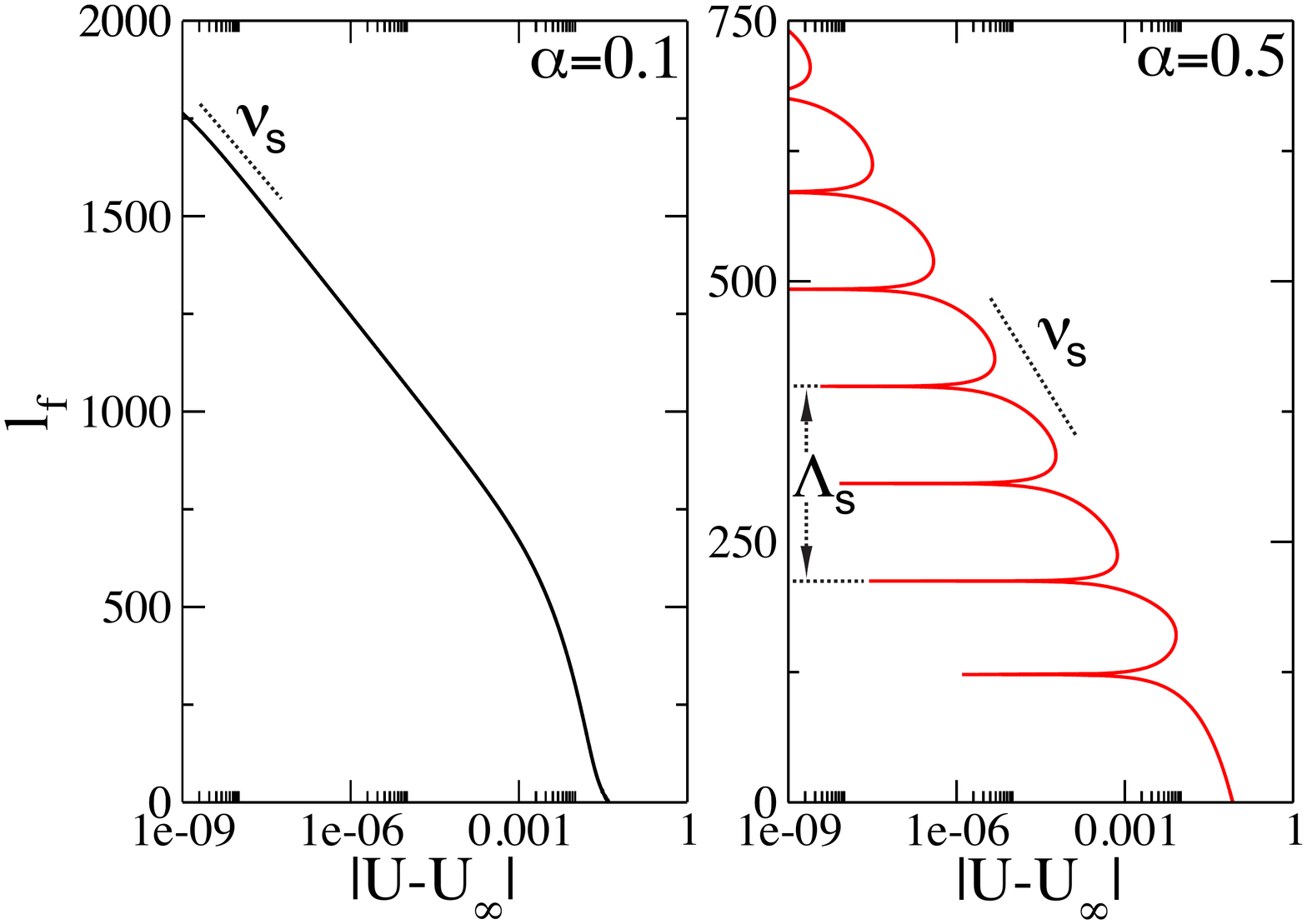}
\caption[Bif]{Comparison of bifurcation diagrams for two inclination
  angles as stated in the panels in th ecase without temperature
  gradient ($\Omega=0$). Top: Left panel: Asymptotic monotonic
  increase of the foot length  $\l_f$ towards the vertical asymptote
  at $U=U_{\infty}$ as a function of the plate velocity $U$ for
  $\alpha=0.1$, which is below $\alpha_c$. Right panel: Snaking
  behaviour of the foot length $\l_f$ where the bifurcation curve
  oscillates around a vertical asymptote at $U=U_\infty$ with decaying
  amplitude of oscillations as a function of the plate velocity $U$
  for $\alpha=0.5$, which is above $\alpha_c$. Note the appearance of
  pairs of saddle nodes (the first being at $U_b$) where the system successively switches branches and ``snakes'' around $U_{\infty}$. Bottom: In order to illustrate the different behaviour for angles below and above $\alpha_c$, we show the  foot-length measure $\l_f$ versus $|U-U_{\infty}|$ in a semi-log plot. Left panel: The semi-log plot shows an asymptotic monotonic growth in $U$. Right panel: An exponential -- oscillating periodic decay is clearly shown. A periodic structure with a snaking wavelength $\Lambda_s$ and an exponential decay rate $\nu_s$ appears after $U_b$ (bifurcation: appearance of the first saddle node).} 
\label{fig:FIG2}
\end{figure}

\section{Numerical results}

In this section, we present numerical solutions of eq.~(\ref{eq:steady_eq}). 
We solve the equation on the domain $[-L_1,\,L_2]$. At $x=-L_1$, we impose the boundary conditions $h'(-L_1)=0$ and $h''(-L_1)=0$. At $x=L_2$, we impose the boundary condition obtained by truncating the asymptotic expansion (\ref{eq:h_4}) for $\Omega\neq 0$ or (\ref{eq:ansatz}) for $\Omega= 0$ and evaluating it at $x=L_2$. 
We additionally impose a condition for the derivative of $h$ at $L_2$ obtained by differentiating the asymptotic expansion for $h$ and evaluating it at $x=L_2$.
To solve this boundary-value problem numerically, we use the
continuation and bifurcation software AUTO-07p (see
refs.~\cite{DKK91,DKK91b}). A description of the application of
numerical continuation techniques to thin film problems can be found in
sect.~4b of the review in ref.~\cite{Dijk13}, in sect.~2.10 of
ref.~\cite{Thie07}, and in refs.~\cite{TBBB03,BeTh10,TBT13}. We perform our numerical calculations on a domain with $L_1=9800$ and $L_2=200$ and choose $G=0.001$.

In fig.~\ref{fig:FIG3}, we compare the numerical solutions with the derived asymptotic expressions for $h$ as $x\rightarrow\infty$, when the inclination angle is $\alpha=0.5$. In the left panel, $\Omega=0$ and $U=0.084$. The solid line shows a numerically computed profile, in which we can identify three regions, namely, a thin precursor film, a foot, and a bath region. We also show the truncated asymptotic expansion (\ref{eq:ansatz}) with one, two, three and four terms included, as is indicated in the legend. In the right panel, $\Omega=0.001$ and $U=0.076$. The solid line shows a numerically computed profile the remaining lines correspond to the truncated asymptotic expansion (\ref{eq:h_4}) with one, two, three, four and five terms included, as is indicated in the legend. In both cases, we can observe that the numerically computed profiles agree with the derived asymptotic expansions and including more terms gives better agreement. 

In fig.~\ref{fig:FIG2}, we present bifurcation diagrams showing the
dependence of a certain solution measure quantifying the foot length
on the velocity of the plate for $\Omega=0$. More precisely,
  the measure is defined by $\l_{f} = (V-V_0)/(h_f-h_p)$, where
  $V=\int_{-L_1}^{L_2}(h(x)-h_p)\mathrm{d}x$, $h_f$ is the
  characteristic foot height, $h_p$ is the precursor film height for
  the corresponding velocity, and $V_0$ is equal to $V$ computed at
  $U=0$.
 
We observe that there is a critical inclination angle, $\alpha_c\approx0.1025$, such that for $\alpha<\alpha_c$, the bifurcation curve increases monotonically towards a vertical asymptote at some value of the velocity, which we denote by $U_\infty$. This can be observed in the left panels of fig.~\ref{fig:FIG2} when $\alpha=0.1$. When $\alpha>\alpha_c$, we observe a snaking behaviour where the bifurcation curve oscillates around a vertical asymptote at $U=U_\infty$ with decaying amplitude of oscillations. This can be observed in the right panels of fig.~\ref{fig:FIG2} when $\alpha=0.5$. We note that in this case there is an infinite but countable number of saddle-nodes at which the slope of the bifurcation curve is vertical.
 
Note that $U_{\infty}$ is different for each inclination
angle. {The character of the steady solutions is discussed below
at fig.~\ref{fig:FIG4B}.}

\begin{figure*}
\includegraphics[width=0.49\hsize]{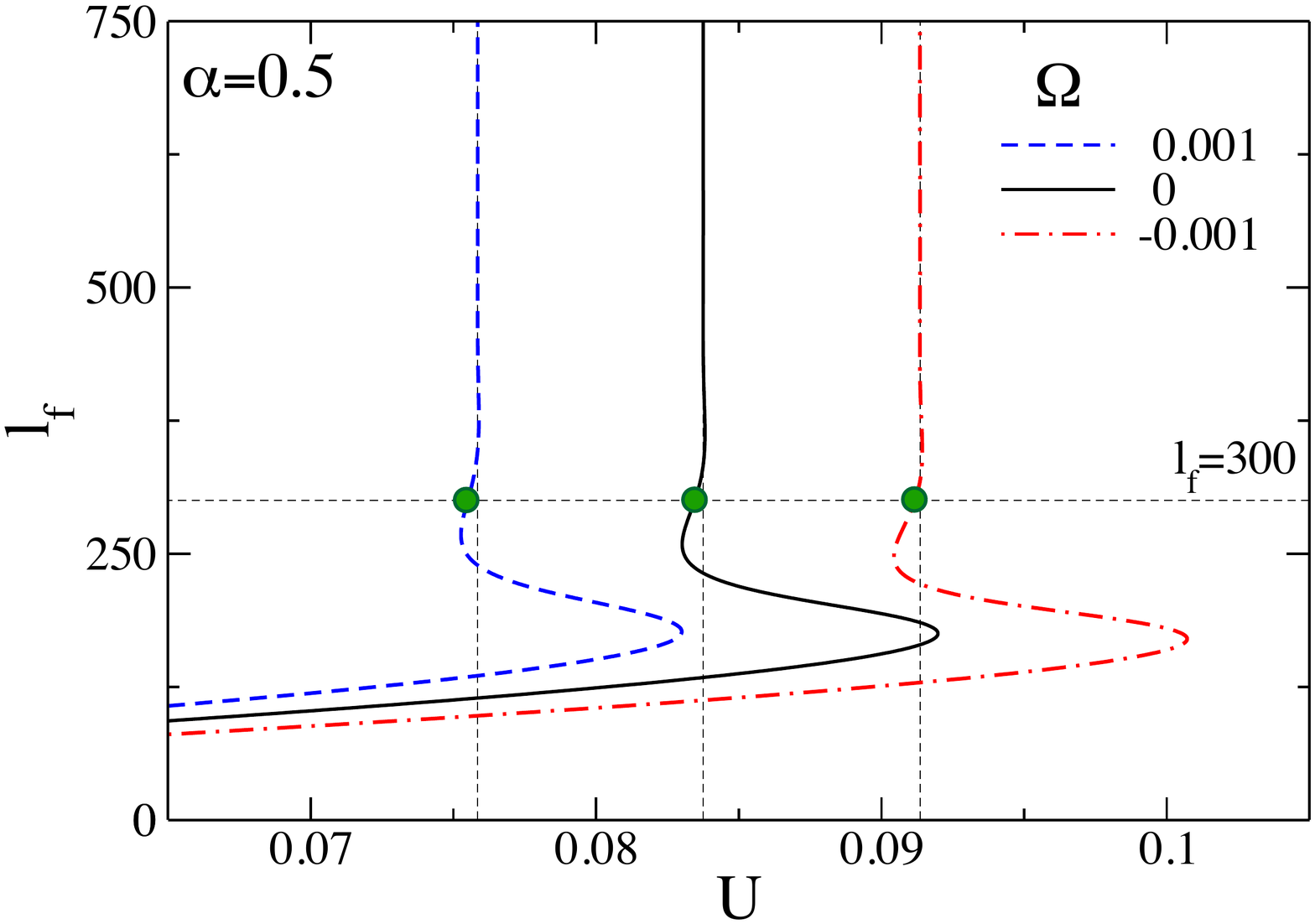}
\includegraphics[width=0.49\hsize]{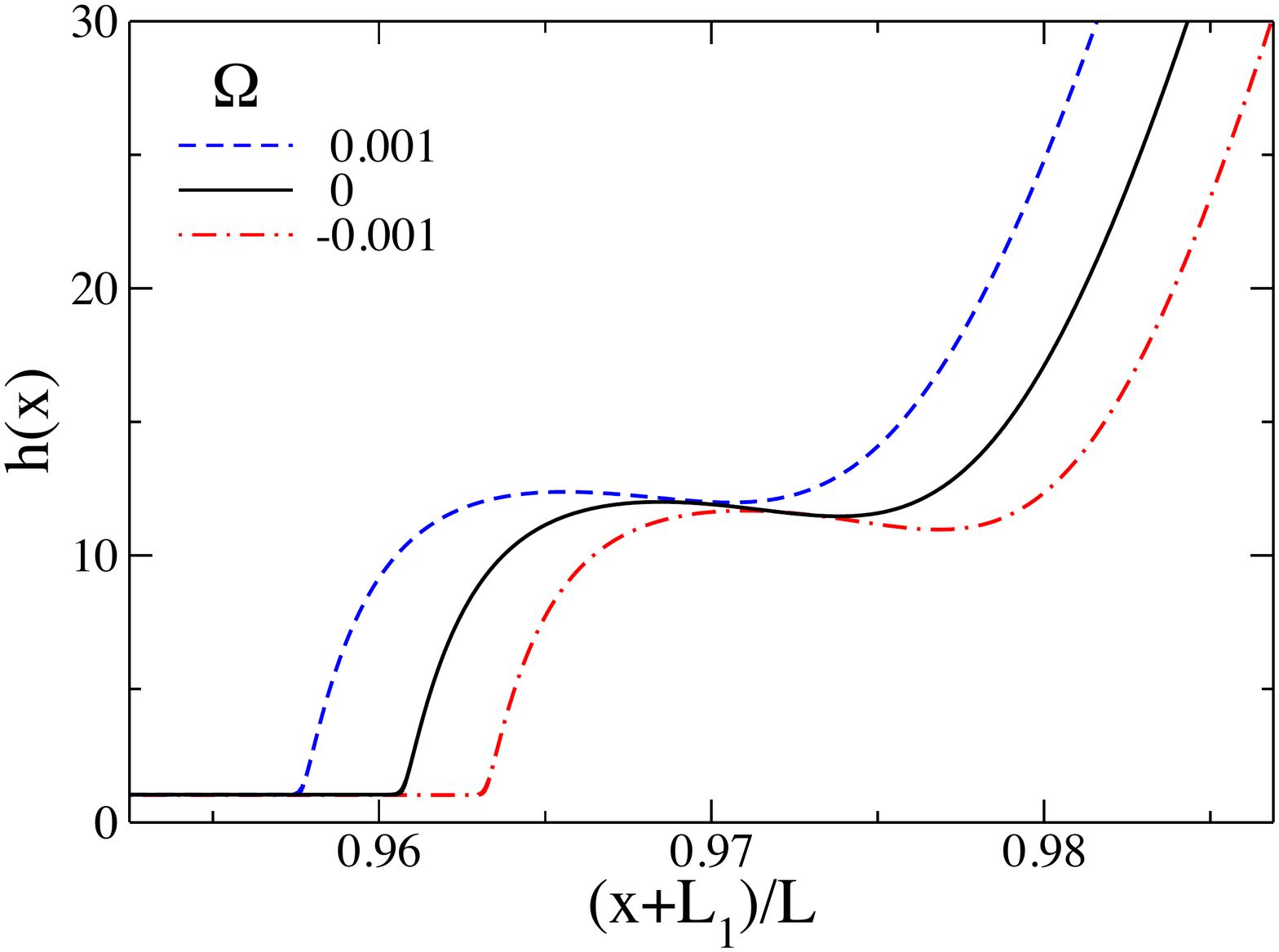}
\caption[$\Omega\neq0$]{Left panel compares bifurcation diagrams for different temperature gradients $\Omega$ as shown in the legend for an inclination angle $\alpha=0.5$. The green filled
circles indicate the points at which $\l_f=300$ and the corresponding film profiles are shown in the right panel. 
Note that the snaking behaviour is present. The temperature gradient $\Omega$ shifts the vertical asymptote at $U_\infty$ and changes the characteristic foot height at $U_\infty$.}
\label{fig:FIG4C}
\end{figure*}

\begin{figure*}
\includegraphics[width=0.49\hsize]{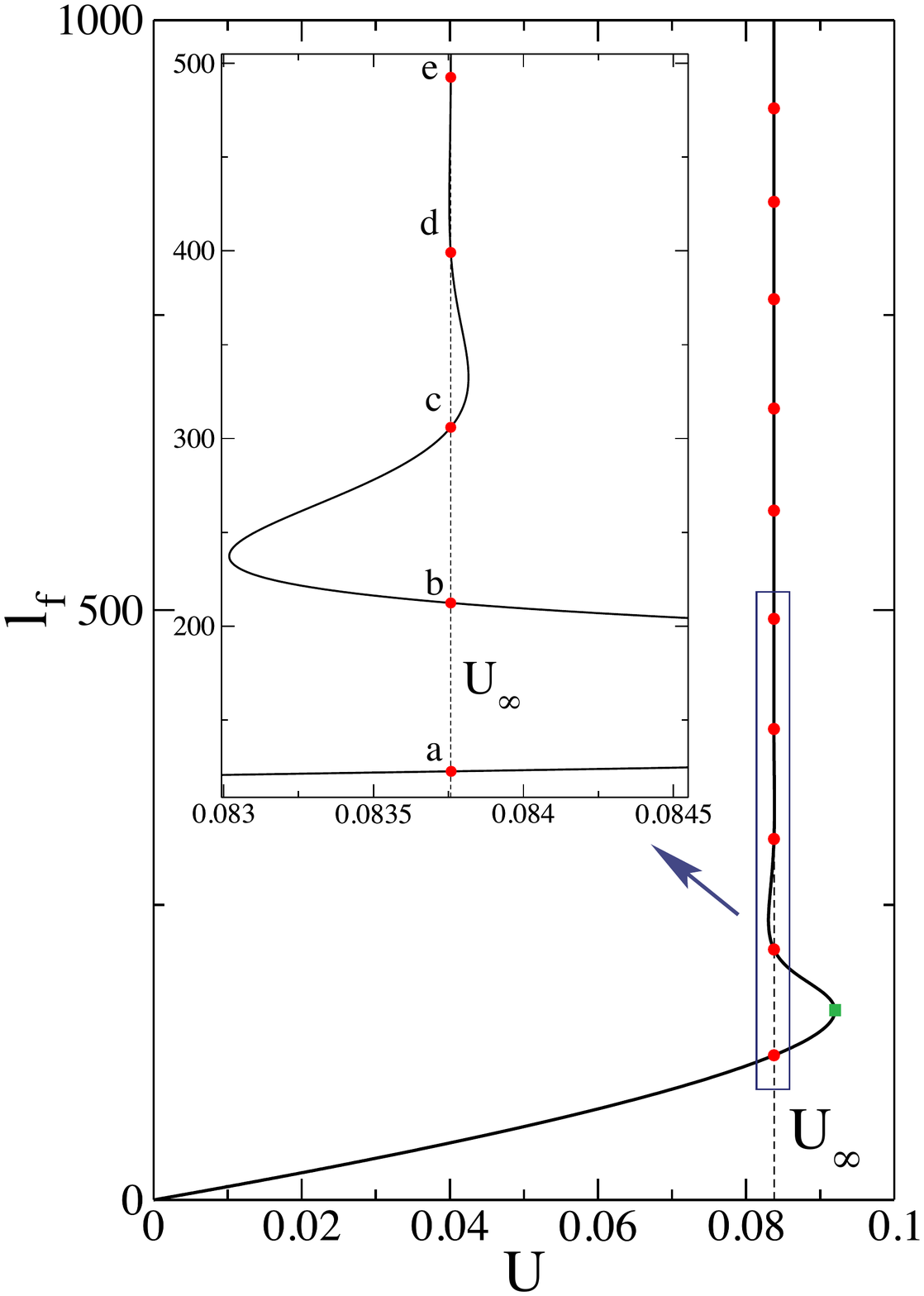}
\includegraphics[width=0.49\hsize]{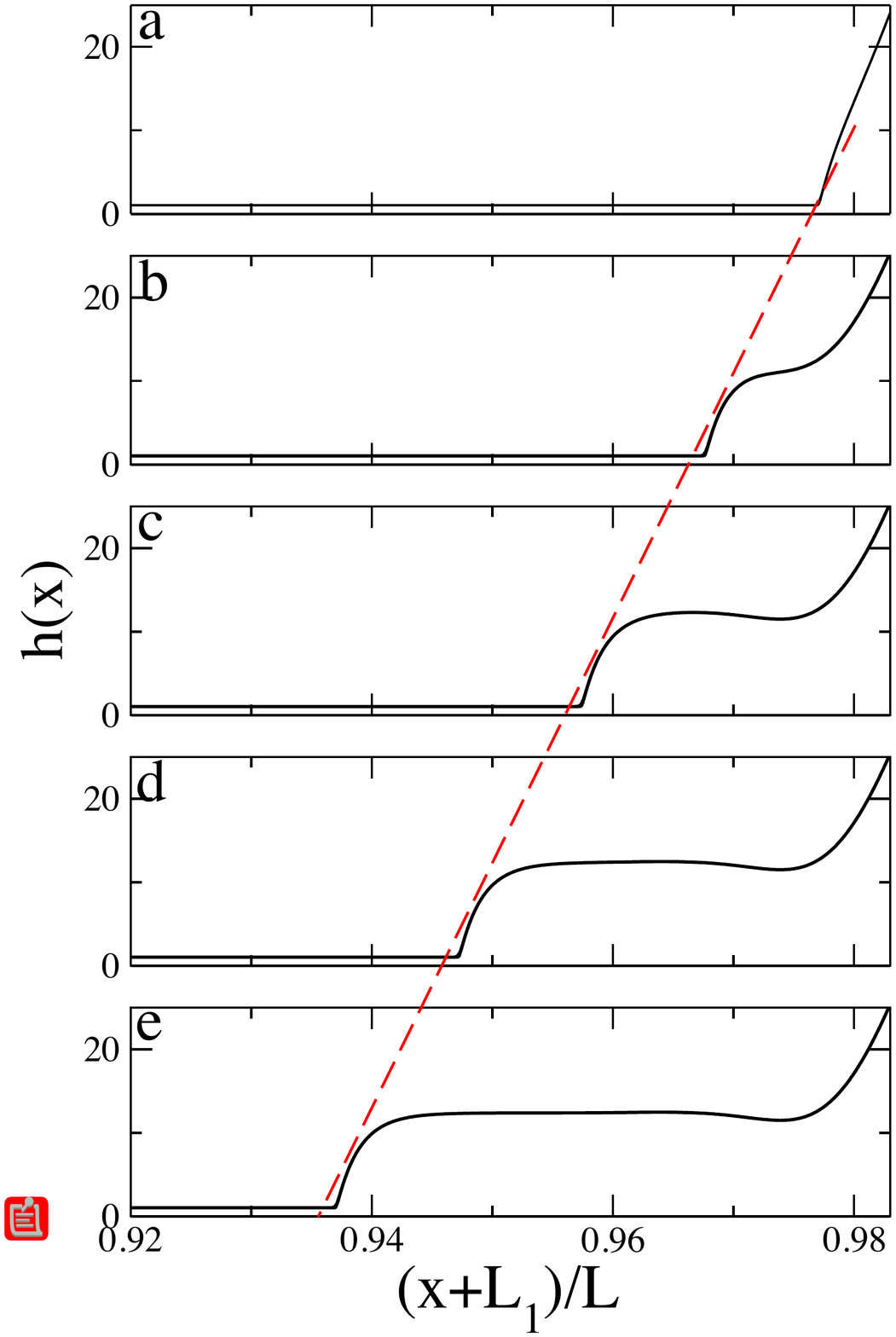}
\caption[Profiles]{Film profiles at plate velocity $U_\infty$ for
  $\alpha = 0.5$. Left panel: Bifurcation diagram. The red filled
  circles correspond to film solutions at plate velocity
  $U_\infty$. The inset shows a blow-up of the region with the first
  five solutions. Note the appearance of a characteristic
    snaking behaviour around $U_\infty$. The letters in the inset
  correspond to the film profiles depicted in the right panel. Note
  the appearance of undulations on top the foot-like part of the
  solution as the foot becomes longer. The numerical domain size used
is $L=10000$, $L_1=9800$. Note that the first profile (a)
  corresponds to a meniscus solution. It is located on the lowest
  branch before the bifurcation curve folds back at $U_b$ (the green
  square). The red dashed line indicates a linear increase in foot length.}
\label{fig:FIG4}
\end{figure*}

We note that in the case with an additional temperature gradient ($\Omega\neq 0$) we observe qualitatively similar bifurcation diagrams. If an inclination angle is below a critical value (which now depends on $\Omega$), then the bifurcation diagrams are monotonic. Otherwise, the bifurcation diagrams show snaking behaviour, as for the case of zero temperature gradient. An example of snaking bifurcation curves for $\alpha=0.5$ and $\Omega=-0.001,\,0$ and $0.001$ is given in fig.~\ref{fig:FIG4C}, and the corresponding bifurcation curves are shown by dashed, solid and dot-dashed lines. We can observe that as the temperature-gradient parameter $\Omega$ is increased/decreased, the vertical asymptote is shifted to the left/right. We can also conclude that if the temperature gradient pulls the liquid downwards, steady-state solutions of this bifurcation branch exist for larger values of $U$. Otherwise, if the temperature gradient pulls the liquid upwards, steady-state solutions of this bifurcation branch exist for smaller values of $U$. The right panel of fig.~\ref{fig:FIG4C} shows three profiles for $l_f=300$ by dashed, solid and dot-dashed lines for $\Omega=0.001,\, 0$ and $-0.001$, respectively.  We observe that the foot height decreases as $\Omega$ decreases.

In order to illustrate the different behaviour for angles below and above $\alpha_c$, we also show the foot length measure, $\l_{f}$, versus $|U-U_{\infty}|$ in a semi-log plot, see the lower left and right panels of fig.~\ref{fig:FIG2} for $\alpha=0.1$ and $\alpha=0.5$, respectively. For $\alpha=0.1$, it can be clearly seen that the bifurcation curve approaches the vertical asymptote exponentially with a rate which we denote by $\nu_s$.
For $\alpha=0.5$, we can see that the approach of the vertical asymptote is exponential with the snaking wavelength tending to a constant value, which we denote by $\Lambda_s$.

Figure~\ref{fig:FIG4} shows the identified snaking behaviour for $\alpha=0.5$ in more detail. 
In the left panel, we see the bifurcation diagram where the red filled
circles correspond to solutions at $U_\infty$. In the chosen solution
measure, the solutions appear equidistantly distributed. In the inset,
the first five solutions are indicated and labeled by (a)-(e) and the
corresponding film profiles are shown in the right panel. The dashed
line in the right panel confirms the linear growth of the foot
length. 

\begin{figure*}
\begin{center}
\includegraphics[width=0.45\hsize]{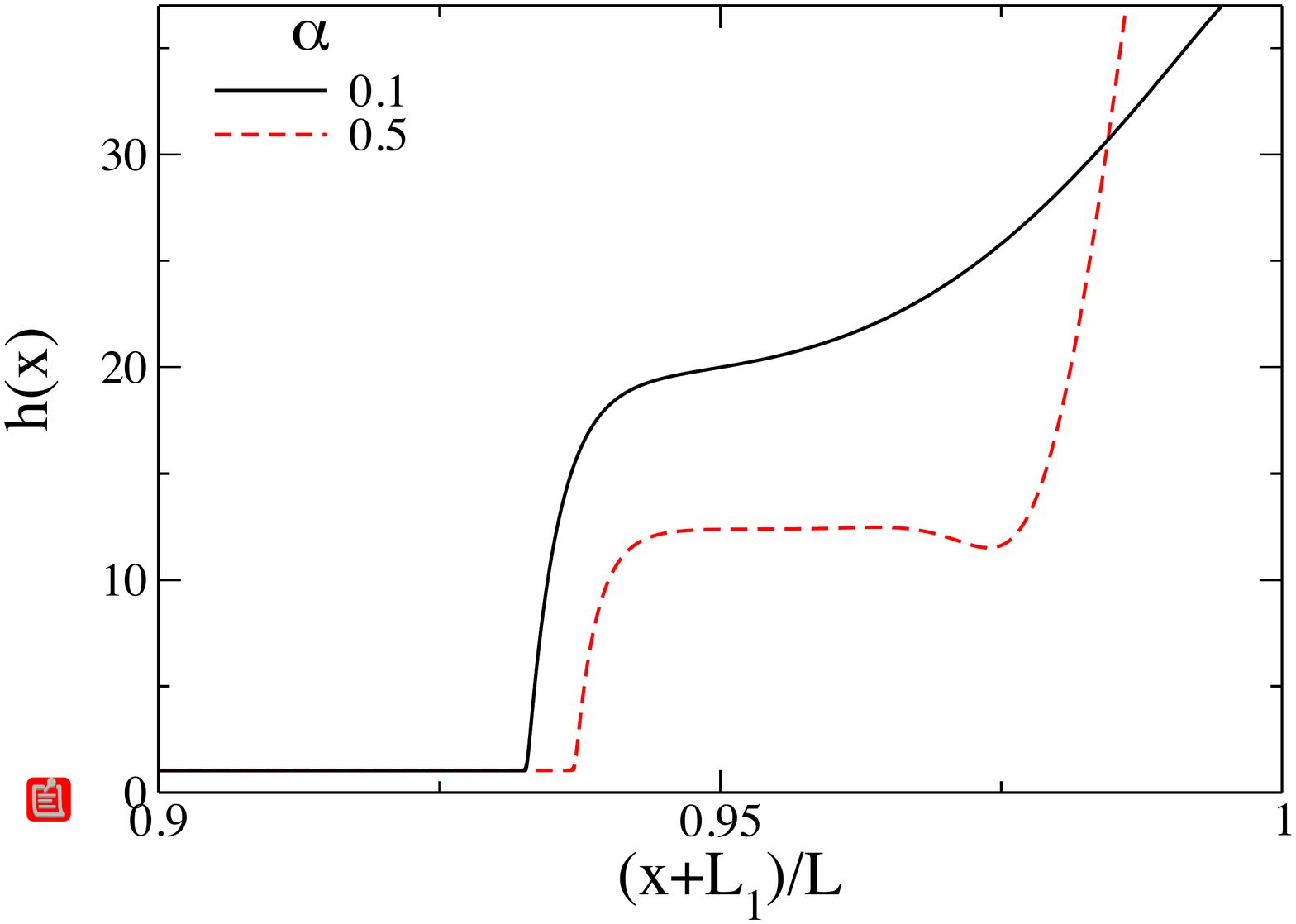}
\includegraphics[width=0.45\hsize]{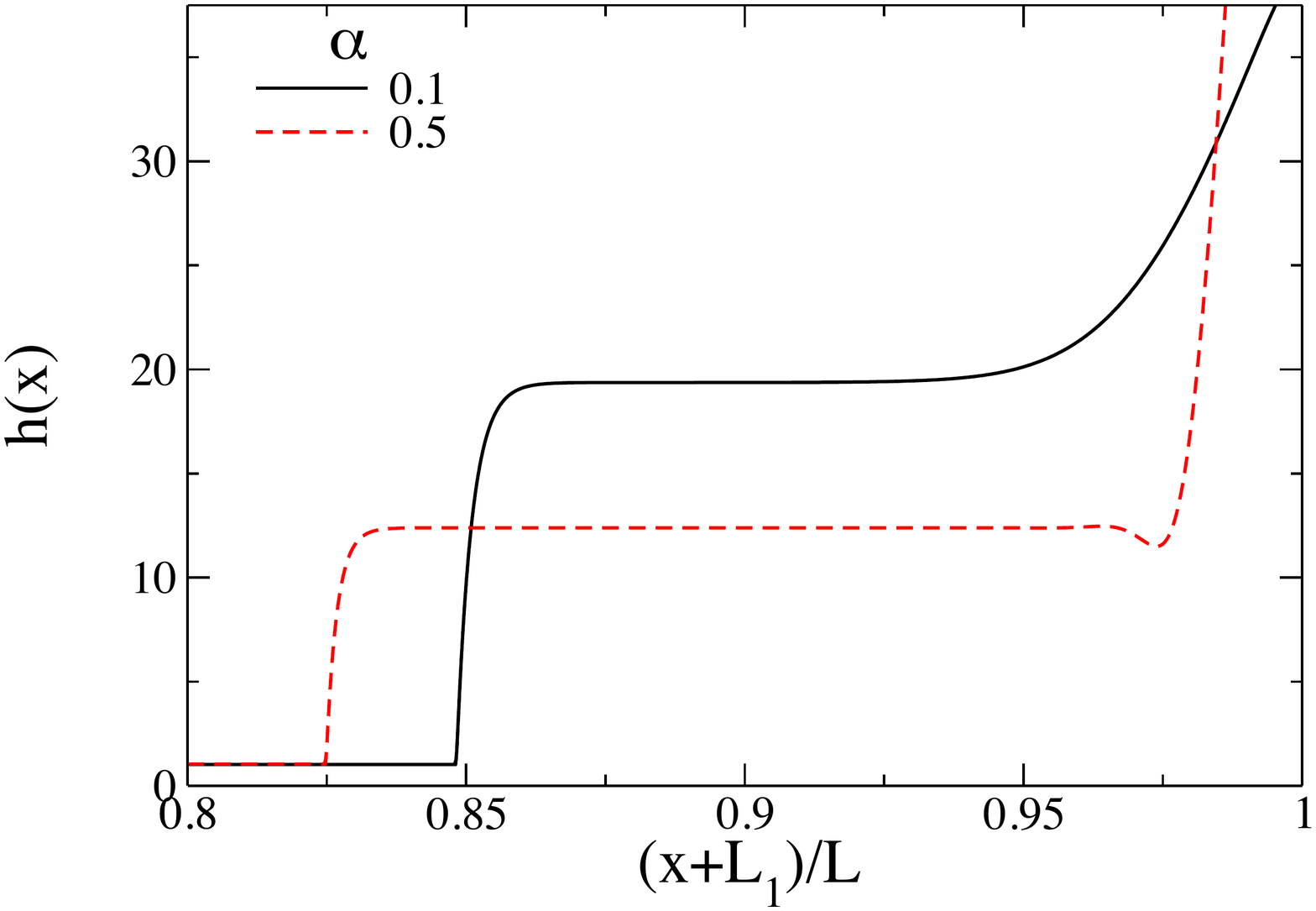}
\includegraphics[width=0.45\hsize]{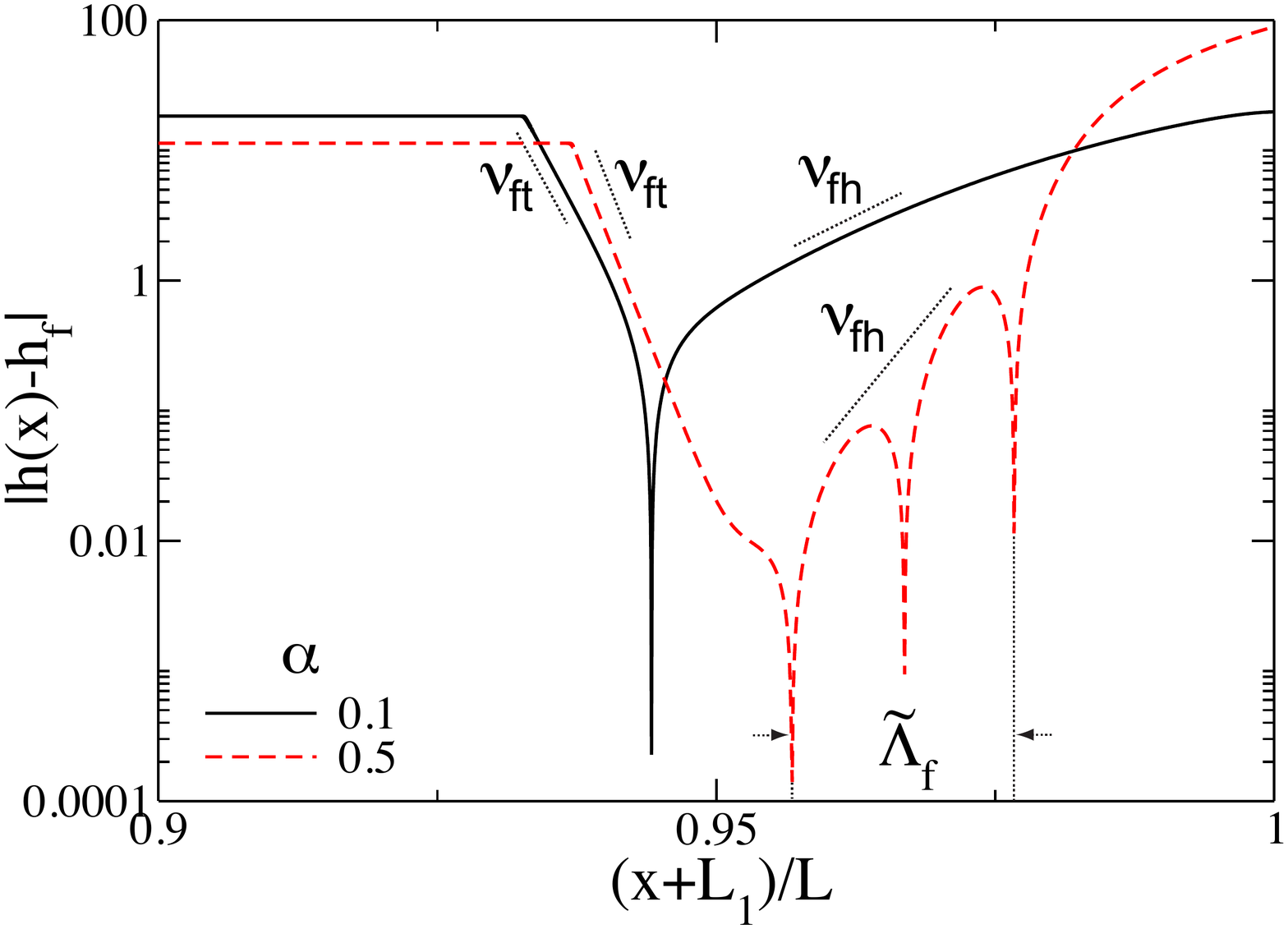}
\includegraphics[width=0.45\hsize]{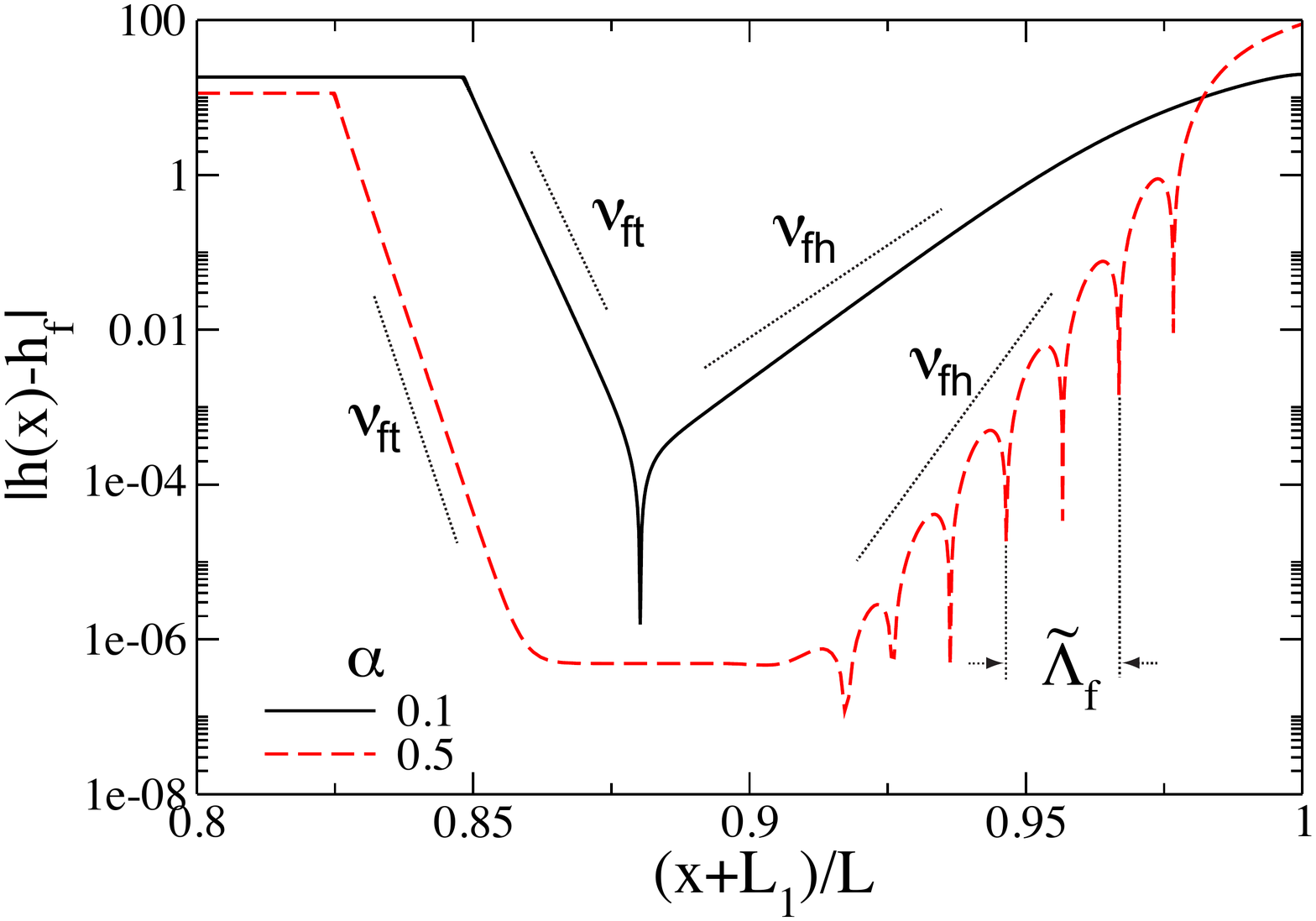}
\caption[Profiles at fixed $U_\infty$ for $\alpha=0.5$]{Film profiles
  below and above $\alpha_c$ given as solid and dashed lines,
  respectively. Left panel: Shown are film profiles for $\alpha=0.1$
  close to $U_\infty$ and for $\alpha=0.5$ at $U_\infty$. Right panel:
  In order to show the appearance of undulations on top of the foot
  above $\alpha_c$, we represent in bottom panels $|h(x)-h_f|$ versus $(x+L_1)/L$ in a
  semi-log plot, where $L_1=9800$, $L=10000$ is the numerical domain
  size and $h_f$ is the characteristic foot height calculated
    for each inclination angle $\alpha$ by solving eq.~(\ref{eq:steady_eq})
    for $h'=0$, $h''=0$ and $h'''=0$  (using the numerically obtained value of the flux
    $C_0$). Observe the exponential approach with rate
    $\nu_\mathrm{fh}$ of the
    foot height from the bath side, and as well the exponential
    departure with rate $\nu_\mathrm{ft}$ from the foot height towards
    the precursor film (see main text for details). Note that the measured foot wavelength is  $\Lambda_{f}=\widetilde{\Lambda}_{f} L$. }
\label{fig:FIG4B}
\end{center}
\end{figure*}

The differences in film profiles for angles below and above $\alpha_c$ can be seen in fig.~\ref{fig:FIG4B} that shows solutions for velocities close to $U_\infty$ for $\alpha=0.1$ and at $U_\infty$ for $0.5$ by solid and dashed lines, respectively.
In the left and the right panels, we compare short-foot and long-foot solutions, respectively, with similar foot lengths. To emphasise the differences, we represent the profiles in a semi-log plot $|h(x)-h_f|$ versus $(x+L_1)/L$ in the bottom panels. For $\alpha=0.1$ we see no undulations -- only exponential decays at a rate denoted by $\nu_\mathrm{fh}$ from the bath to the foot and at a rate denoted by $\nu_\mathrm{ft}$ from the foot to the precursor.  However, for $\alpha=0.5$ we observe an oscillatory exponentially decaying behaviour at a rate denoted by $\nu_\mathrm{fh}$ with a wavelength denoted by $\Lambda_{f}$ in the region between the bath to the foot.  In the region between the foot and the precursor film, we again observe an exponential decay. 

{Figures~\ref{fig:FIG2} to~\ref{fig:FIG4} allow us to recognise the
  observed behaviour as collapsed heteroclinic snaking
  \cite{noteSnaking}: The bifurcation curve in fig.~\ref{fig:FIG4} is
  a snaking curve of heteroclinic orbits, {\it i.e.}, each point on the
  curve represents a heteroclinic orbit connecting the fixed points for
  precursor film $\zh{y}_p$ and bath surface $\zh{y}_b$ of the
  dynamical system (\ref{eq:y1_prime})-(\ref{eq:y3_prime}), namely, if
  $h_p$ and $h_f$ are the heights of the precursor film and the foot
  and $\alpha$ is the inclination angle, then the fixed points are
  $\zh{y}_p=(1/h_p,\,0,\,0)$ and $\zh{y}_b=(0,\,\alpha,\,0)$,
  respectively. In the limit $U\to U_\infty$ the curve approaches a
  heteroclinc chain consisting of two heteroclinc orbits -- one
  connecting the fixed points precursor film $\zh{y}_p$ and foot film
  $\zh{y}_f=(1/h_f,\,0,\,0)$ and the other one connecting foot
  $\zh{y}_f$ and bath $\zh{y}_b$. Figure~\ref{fig:FIG4} (left) shows the
  first 5 heteroclinic orbits connecting $\zh{y}_p$ and $\zh{y}_b$ -- all at
  $U=U_\infty$. In sect.~5 it is proved that at $U=U_\infty$ there
  exists a countable infinite number of such heteroclinic
  connections.}

\begin{figure}
\vspace{-1cm}
\begin{center}
\includegraphics[width=1\hsize]{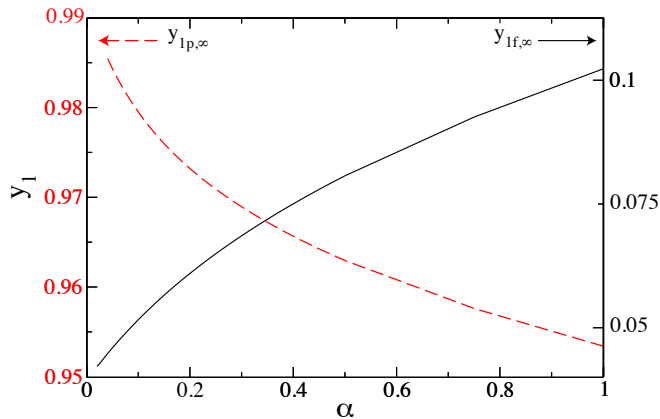}
\vspace{-0.5cm}
\caption[$y_1$ at precursor and foot fixed points]{$y_1$ ($y_1=1/h$)
  at fixed points for precursor film height, $h_p$, ($y_{1p}=1/h_p$),
  and foot film height, $h_f$, ($y_{1f}=1/h_f$), versus inclination
  angle $\alpha$  at $U_{\infty}$ shown by dashed and solid lines,
  respectively, in a double entry plot. Note that the correct numerically
  obtained flux $C_0$ is needed at each $\alpha$ to determine the
  fixed points.
The left side of the ordinate axis corresponds to the precursor film, the right side corresponds to the foot.}
\label{fig:y1}
\end{center}
\end{figure}

\begin{figure}
\begin{center}
\includegraphics[width=1\hsize]{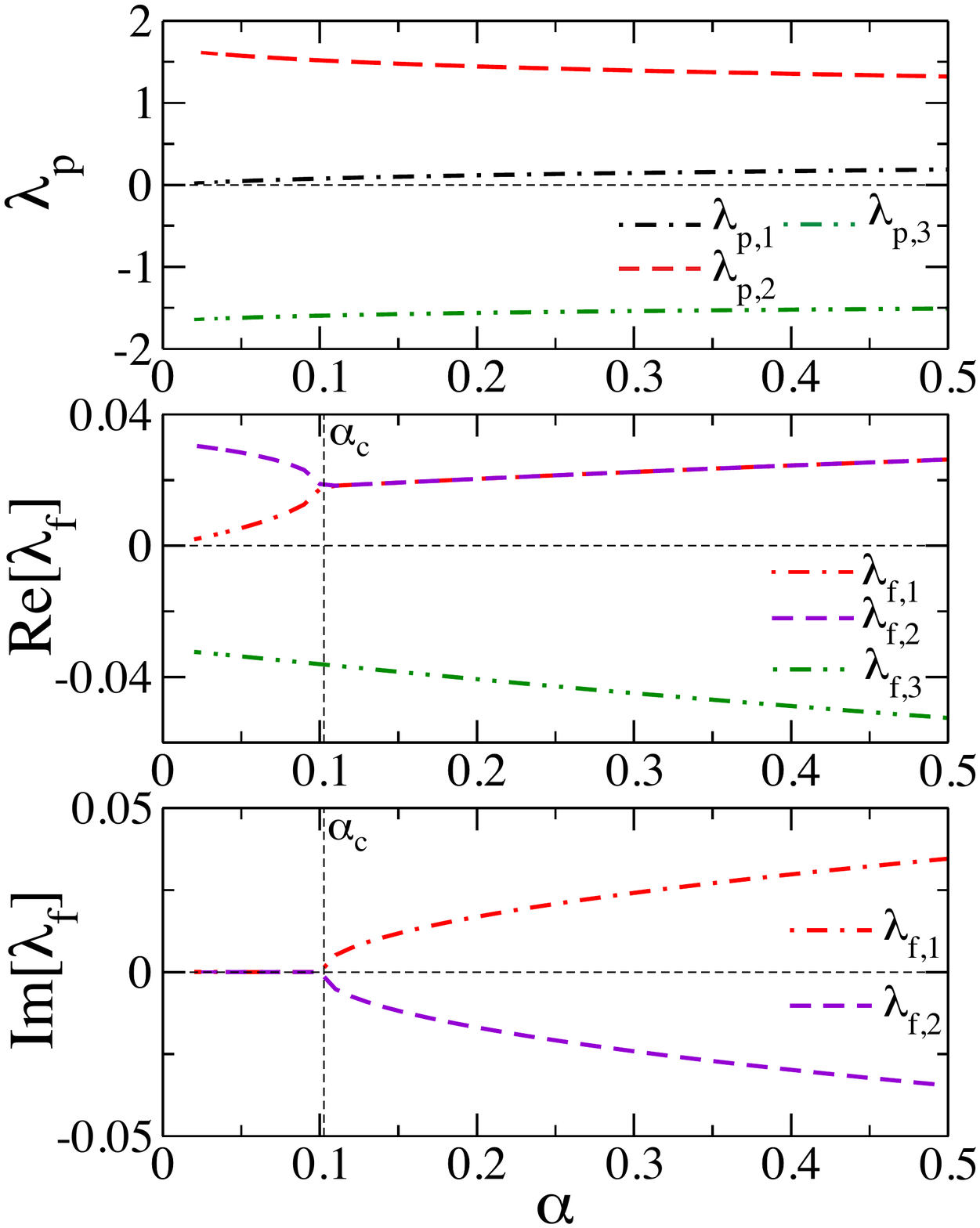}
\vspace{-1cm}
\caption[Eigenvalues]{Eigenvalues at corresponding $U_\infty$ for each
  $\alpha$. Upper panel: Shown are the three eigenvalues
  $\lambda_p$  versus $\alpha$ for the fixed point corresponding to the precursor film. Note that all the eigenvalues are real. 
  Middle and bottom panels:  Show are the real and the imaginary parts, respectively, of the three eigenvalues $\lambda_f$  versus
   $\alpha$ for the fixed corresponding to the foot.}
\label{fig:FIG5}
\end{center}
\end{figure}

The values of $h_p$ and $h_f$ at $U=U_\infty$ are shown in
fig.~\ref{fig:y1} as functions of $\alpha$ by dashed and solid lines,
respectively. In fig.~\ref{fig:FIG5}, we show the dependence of the
eigenvalues of the Jacobians of system
(\ref{eq:y1_prime})-(\ref{eq:y3_prime}) at fixed points $\zh{y}_p$ and
$\zh{y}_f$ at $U=U_\infty$ as functions of $\alpha$ {(also
  cf.~beginning of sect.~5).} We note that for the precursor film
all the eigenvalues are real, two of them are positive and one is
negative independently of the angle.  We denote these eigenvalues by
$\lambda_{p,i}$,
$i=1,\,2,\,3$. 
However for the foot, the behaviour of the eigenvalues changes for
angles below and above a critical value and it turns out that this
critical angle is the same as the critical angle at which monotonic
bifurcation diagrams change to snaking, {\it i.e.},
$\alpha_c\approx0.1025$. We observe that for $\alpha<\alpha_c$ all the
eigenvalues for the foot are real -- two are positive and denoted by
$\lambda_{f,1}$ and $\lambda_{f,2}$ so that
$\lambda_{f,1}<\lambda_{f,2}$ and one is negative and is denoted by
$\lambda_{f,3}$. However, for $\alpha>\alpha_c$ there is a negative
real eigenvalue, $\lambda_{f,3}$, and a pair of complex conjugate
eigenvalues with positive real parts, $\lambda_{f,1}$ and
$\lambda_{f,2}$. Table~\ref{tab:1} shows the values of eigenvalues
$\lambda_{f,i}$, $i=1,\,2,\,3$, for $\alpha=0.1$ and $0.5$.

\rowcolors{1}{lightgray}{}{}
\begin{table}[h!]
    \caption[Eigenvalues] {Eigenvalues at fixed point  
      $\zh{y}_{f}=(y_{1f},\,0,\,0)$ with $y_{1f}=1/h_f$ 
      for $\alpha=0.1$ close to $U_\infty$ and for $\alpha=0.5$ at
      $U_\infty$. Note that all the eigenvalues are real for $\alpha=0.1$, whereas for $\alpha=0.5$ one eigenvalue is real and negative and two are complex conjugates with positive real parts.
See fig.~\ref{fig:FIG5}.} 
\label{tab:1}
\centering
\begin{tabular}{|c|c|c|l|l|c|}
\hline
\textbf{$\alpha$} & $h_f$           & $y_{1f}$ & $\lambda_{f,1}$             & $\lambda_{f,2}$     & $\lambda_{f,3}$             \\ \hline
\rowcolor{white}    0.1     &     19.3732 &  0.0516      & 0.0173            &    0.0188 &    -0.0361         \\ \hline
\rowcolor{white}   0.5      &     12.3922 &  0.0807     &0.0263                    & 0.0263                      & -0.0525 \\ 
\rowcolor{white}             &                    &                  &$+\mathrm{i}\: 0.0346$ & $-\mathrm{i}\: 0.0346$    &  \\ \hline
\end{tabular}
\end{table}

\rowcolors{1}{lightgray}{}{}
\begin{table}
\centering
\caption[]{ Shown is the comparison of the exponential decays
  $\nu_\mathrm{ft}$, $\nu_\mathrm{fh}$ with the eigenvalue $\nu$ from
  the linear stability analysis for $\alpha=0.1$ close to $U_\infty$
  and for $\alpha=0.5$ at $U_{\infty}$ for solutions with a short
  foot. See fig.~\ref{fig:FIG4B}.}
\label{tab:4}
\begin{tabular}{|c|c|c|c|c|}
\hline
\textbf{$\alpha$} &$\nu=\mathrm{Re}[\lambda_{f,3}]$ & $\nu_\mathrm{ft}$    &$\nu=\mathrm{Re}[\lambda_{f,1}]$ &$\nu_\mathrm{fh}$\\ \hline
\rowcolor{white}    0.1      &   -0.0361    & -0.0403            & 0.0173     & 0.0152 \\ \hline
\rowcolor{white}    0.5      &-0.0525&-0.0497 & 0.0263 & 0.0278 \\ \hline
\end{tabular}
\end{table}
\rowcolors{1}{lightgray}{}{}
\begin{table}[h!]
\centering
\caption[]{Shown is the comparison of the exponential decays
  $\nu_\mathrm{ft}$, $\nu_\mathrm{fh}$ with the eigenvalue $\nu$ from
  the linear stability analysis for $\alpha=0.1$ close to $U_\infty$
  and for $\alpha=0.5$ at $U_{\infty}$ for solutions with a long foot. See fig.~\ref{fig:FIG4B}.}
\label{tab:5}
\begin{tabular}{|c|c|c|c|c|}
\hline
\textbf{$\alpha$} &$\nu=\mathrm{Re}[\lambda_{f,3}]$ & $\nu_\mathrm{ft}$    &$\nu=\mathrm{Re}[\lambda_{f,1}]$ &$\nu_\mathrm{fh}$\\ \hline
\rowcolor{white}    0.1      &   -0.0361      &  -0.0356             & 0.0173   & 0.0155\\ \hline
\rowcolor{white}    0.5      &-0.0525& -0.0463 & 0.0263& 0.0255\\ \hline
\end{tabular}
\end{table}

\rowcolors{1}{lightgray}{}{}
\begin{table}
\centering
\caption[]{Shown is the comparison of the wavelength of  snaking
  $\Lambda_s$ from the bifurcation diagram and wavelength of the undulations of the foot $\Lambda_f$  from the foot-like profile with the wavelength $\Lambda$ calculated from the eigenvalues $\lambda_{f,i}$  at $U_{\infty}$ for  $\alpha=0.5$. Note the locking between $\Lambda \approx \Lambda_s \approx \Lambda_f$. See fig.~\ref{fig:FIG2} and fig.~\ref{fig:FIG4B}.}
\label{tab:2}
\begin{tabular}{|c|c|c|c|c|}
\hline
\textbf{$\alpha$} & $\Lambda\!\!=\!\!2\pi /\mathrm{Im}[\lambda_{f,1}]$&    $\Lambda_f\,$(long) &  $\Lambda_f\,$(short)&  $\Lambda_s$ \\ \hline
\rowcolor{white}   0.5     &    181.6987 &   202.6920   & 198.8801 &184.7657\\ \hline
\end{tabular}
\end{table}

In tables~\ref{tab:4} and \ref{tab:5}, we compare
$\mathrm{Re}[\lambda_{f,3}]$ with the exponential rate
$\nu_\mathrm{ft}$ characterising the connection between the foot and
the precursor film, and $\mathrm{Re}[\lambda_{f,1}]$ with the
exponential rate $\nu_\mathrm{fh}$ characterising the connection
between the foot and the bath. Table~\ref{tab:4} corresponds to a short
foot, while table~\ref{tab:5} corresponds to a long foot. For
$\alpha=0.5$ the plate velocity is equal to $U_\infty$, while for
$\alpha=0.1$ we choose a foot of approximately the same lengths as for $\alpha=0.5$ and we note that for $\alpha=0.1$ the bifurcation curves do not reach $U_\infty$, but for the chosen foot the velocities coincide with $U_\infty$ up to at least seven significant digits. The results show that there is good agreement between $\mathrm{Re}[\lambda_{f,3}]$ and $\nu_\mathrm{ft}$ and between $\mathrm{Re}[\lambda_{f,1}]$ and $\nu_\mathrm{fh}$ for both values of $\alpha$ and for both foot lengths, with a maximal error below $12\%$.

\rowcolors{1}{lightgray}{}{}
\begin{table}
\centering
\caption[]{Shown is the comparison of the exponential decay constant $1/\nu_{S}$ from the bifurcation diagrams with the eigenvalues $\lambda_{f,i}$ calculated from the linear stability analysis for $\alpha=0.1$ and $\alpha=0.5$. See fig.~\ref{fig:FIG2}.}
\label{tab:3}
\begin{tabular}{|c|c|c|c|}
\hline
\textbf{$\alpha$} &$\mathrm{Re}[\lambda_{f,1}]$ & $1/\nu_s$\\ \hline
\rowcolor{white}    0.1      &   0.0173       &0.0151  \\ \hline
\rowcolor{white}    0.5      &0.0263 & 0.0284 \\ \hline
\end{tabular}
\end{table}
\rowcolors{1}{white}{}{}

In table~\ref{tab:2} we compare $\Lambda= 2\pi /\mathrm{Im}[\lambda_{f,1}]$ with the
wavelength of the oscillations on the foot, $\Lambda_f$, for a long
and a short foot, and with the wavelength of oscillations in snaking
bifurcation diagrams, $\Lambda_s$, when $\alpha=0.5$. The results show
that there is good agreement between $\Lambda$ and $\Lambda_s$ -- the
error is below $2\%$, and between $\Lambda$ and $\Lambda_f$ for both
foot lengths -- the error is below $12\%$
{\cite{noteMeasure}}.

In table~\ref{tab:3}, we compare $\mathrm{Re}[\lambda_{f,1}]$ with the exponential rate $1/\nu_s$, where $\nu_s$ is characterises the rate at which the bifurcation diagrams approach the vertical asymptotes. We again observe good agreement for both values of $\alpha$, with  an  error up to $13\%$.

The close agreement between the eigenvalues corresponding to the foot and the quantities obtained from the bifurcation diagrams and the foot profiles is explained in the next section.

\section{{Collapsed} heteroclinic snaking}

In what follows, our aim is to explain the snaking behaviour observed in our numerical results,  see the left panels of fig.~\ref{fig:FIG2} and fig.~\ref{fig:FIG4}. {We perform our analysis in the way similar to the Shilnikov-type method for studying subsidiary homoclinic orbits near the primary one explained in, {\it e.g.}, ref.~\cite{GlSp1984jsp}.} For simplicity, we consider the case of zero temperature gradient, {\it i.e.}, we set $\Omega=0$. First, let us consider fixed points of  system (\ref{eq:y1_prime})-(\ref{eq:y3_prime}) with $y_1\neq 0$. For such fixed points, $y_2=y_3=0$ and $y_1$ satisfies the equation
\begin{equation}
f(y_1)\equiv y_1^3 -\frac{U}{C_0}y_1^2+\frac{G\alpha}{C_0}=0.
\end{equation}
It can be easily checked that this cubic polynomial has a local maximum at $y_1^a=0$ and a local minimum at a  positive point $y_1^b$. Moreover, $f(y_1^a)>0$ implying that there is always a fixed point with a negative value of the $y_1$-coordinate. We disregard this point, since physically it would correspond to negative film thickness. 
Also, assuming that $G\alpha<(4/27)(U^3/C_0^2)$, we obtain $f(y_1^b)<0$, which implies that there are two positive roots $a_1$ and $a_2$ of the cubic polynomial satisfying $a_1<a_2$. This implies that there are two more fixed points, $\zh{y}_f=(a_1,\,0,\,0)$ and $\zh{y}_p=(a_2,\,0,\,0)$. The point $\zh{y}_f$ corresponds to the foot and the point $\zh{y}_p$ corresponds to the precursor film.

To analyse stability of these fixed points, we compute the Jacobian at these points,
\begin{equation}
\!J_{\zh{y}_{f,p}}=\!\left(
\begin{array}{ccc}
0 & -a_{1,2}^2 & 0\\
0 & 0 & 1\\
2Ua_{1,2}-3C_0a_{1,2}^2\,\, & \,\,6a_{1,2}^7-3a_{1,2}^4+G\,\, & 0
\end{array}
\right).\!\!
\end{equation}
A simple calculation shows that for both, $\zh{y}_f$ and $\zh{y}_p$, all the eigenvalues have non-zero real parts implying that these points are hyperbolic. Moreover, both points have two-dimensional unstable manifolds and one-dimensional stable manifolds. Our numerical simulations presented in the previous section show that 
{for the values of the inclination angle $\alpha$ that we have considered,} there exists a value of the plate speed, $U_\infty$, such that in the vicinity of this value there exist steady solutions for which the foot length can be arbitrarily long, see fig.~\ref{fig:FIG2}. {(In fact, we found that this is true if $\alpha$ is smaller than a certain transition value $\alpha_T\approx 2.42$. For larger values of $\alpha$, the solution branches originating from $U=0$ are not anymore characterised by such limiting velocities. In the present manuscript, we do not consider such solutions and assume therefore that $\alpha<\alpha_T$. Other types of solutions will be analysed elsewhere.) } We conclude that at $U=U_\infty$, there exists a heteroclinic chain connecting the fixed points $\zh{y}_p$, $\zh{y}_f$ and $\zh{y}_b$. As was discussed in the previous section, in the top panel of fig.~\ref{fig:FIG5}, we can observe that for point $\zh{y}_p$ all the eigenvalues are real at $U=U_\infty$ implying that this point is a saddle. The two bottom panels of fig.~\ref{fig:FIG5} demonstrate that there is a critical inclination angle $\alpha_c\approx 0.1025$ such that for $\alpha\leq\alpha_c$, all the eigenvalues for $\zh{y}_f$ are real, whereas for $\alpha>\alpha_c$, one eigenvalue is real and negative and there is a pair of complex conjugate eigenvalues with positive real parts. Therefore, for $\alpha\leq\alpha_c$, point $\zh{y}_f$ is a saddle, but for $\alpha>\alpha_c$, it is a saddle-focus. In the following Theorem, we analytically prove that if $\zh{y}_f$ is a saddle-focus, there exists  an infinite but countable number of subsidiary heteroclinic orbits connecting $\zh{y}_p$ and $\zh{y}_b$ that lie in a sufficiently small neighbourhood of the heteroclinic chain connecting $\zh{y}_p$, $\zh{y}_f$ and $\zh{y}_b$. This explains the existence of an infinite but  countable number of steady-state solutions having different foot lengths observed in the previous section, see the left panels of fig.~\ref{fig:FIG2} and fig.~\ref{fig:FIG4}. {Note that an infinite but countable number of solutions has
  also been observed in, {\it e.g.}, ref.~[16] for the case of a
  liquid film rising onto a resting inclined plate driven by Marangoni
  forces due to a temperature gradient. There, the authors identify type
  1 and type 2 solutions with small and large far-field thicknesses,
  respectively. These correspond to our precursor and foot height,
  respectively.  It is observed that for certain parameter values
  there exists an infinite but countable number of type 2
  solutions. Similar to our case, this can be explained
  by the existence of a heteroclinic chain connecting the three fixed
  points. However, unlike here, in ref.~[16] the chain connects the fixed point for
  the thick film along its unstable manifold with the fixed
  point for the thin film thickness that is then connected with the fixed point
  for the bath.}

\begin{figure}
\centering
\includegraphics[width=1\hsize]{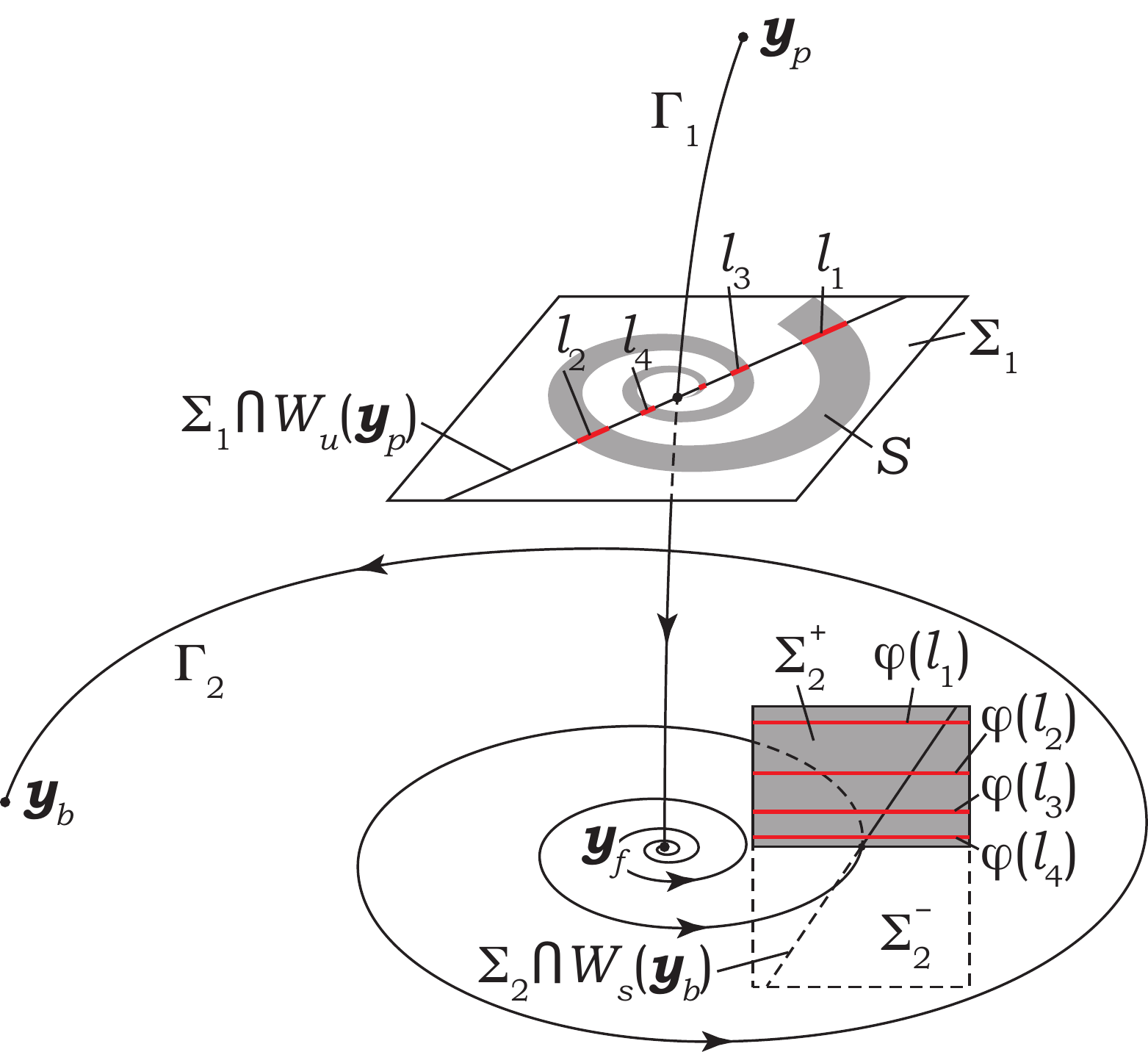}
\caption{Schematic representation in the three-dimensional phase-space of the fixed points $\zh{y}_p$, $\zh{y}_f$ and $\zh{y}_b$ of system (\ref{eq:general_sys}) when $\beta=\beta_0$. 
The fixed point $\zh{y}_p$ is a saddle point with two-dimensional unstable manifold, $W_u(\zh{y}_p)$, and a one-dimensional stable manifold. The fixed point is $\zh{y}_f$ is a saddle-focus  with two-dimensional unstable manifold and a one-dimensional stable manifold. The fixed point $\zh{y}_b$ is a non-hyperbolic point having two-dimensional stable manifold, $W_s(\zh{y}_b)$.    The fixed points $\zh{y}_p$ and $\zh{y}_f$ are connected by the heteroclinic orbit $\Gamma_1$ and the fixed points $\zh{y}_f$ and $\zh{y}_b$ are connected by the heteroclinic orbit $\Gamma_2$. }
\label{fig:phase_space}
\end{figure}

{\bf Theorem.} Consider a three-dimensional system
\begin{equation}
\zh{y}'=\zh{f}(\zh{y},\,\beta),\qquad \zh{y}\in\mathbb{R}^3,\label{eq:general_sys}
\end{equation}
where $\beta$ denotes a parameter {(that takes the role of the plate
velocity $U$ above). }
We assume that there exist three fixed points, which we denote by
$\zh{y}_p$, $\zh{y}_f$ and $\zh{y}_b$, when $\beta$ is sufficiently
close to a number $\beta_0$ {({\it{i.e.}}, a number like the plate
velocity $U_\infty$).} 
We additionally assume that $\zh{y}_p$ and $\zh{y}_b$  
have a two-dimensional unstable manifold $W_u(\zh{y}_p)$ and a
two-dimensional stable manifold $W_s(\zh{y}_b)$, respectively, and
that $\zh{y}_f$ is a saddle-focus fixed point with a one-dimensional
stable manifold $W_s(\zh{y}_f)$ and a two-dimensional unstable
manifold $W_u(\zh{y}_f)$ ({\it i.e.}, the eigenvalues of the Jacobian
at $\zh{y}_f$ are $-\lambda_{1}$, $\lambda_2\pm\mathrm{i}\:\omega$,
where $\lambda_1=\lambda_1(\beta)$, $\lambda_2=\lambda_2(\beta)$ and
$\omega=\omega(\beta)$ are positive real numbers when $\beta$ is
sufficiently close to $\beta_0$) {\cite{noteManifold}}. Let us also assume that for $\beta=\beta_0$, there is a heteroclinic orbit $\Gamma_1\in W_u(\zh{y}_p)\cap W_s(\zh{y}_f)$ connecting $\zh{y}_p$ and $\zh{y}_f$ and that the manifolds $W_u(\zh{y}_f)$ and $W_s(\zh{y}_b)$ intersect transversely so that there is a heteroclinic orbit $\Gamma_2\in W_u(\zh{y}_f)\cap W_s(\zh{y}_b)$ connecting $\zh{y}_f$ and $\zh{y}_b$. Then for $\beta=\beta_0$ there is an infinite but countable number of heteroclinic orbits connecting $\zh{y}_p$ and $\zh{y}_b$ and passing near $\zh{y}_f$. Moreover, the difference in `transition times' from $\zh{y}_p$ to $\zh{y}_b$ tends asymptotically to $\pi/\omega$ (the meaning of a `transition time' from $\zh{y}_p$ to $\zh{y}_b$ will be explained below).

{\bf Proof:}  
After a suitable change of variables, the dynamical system $\zh{y}'=\zh{f}(\zh{y},\,\beta)$ can be written in the form
\begin{eqnarray}
y_1'&=&\lambda_2 y_1-\omega y_2 +\tilde{f}_1(\zh{y},\,\beta),\\
y_2'&=&\omega y_1+\lambda_2 y_2 +\tilde{f}_2(\zh{y},\,\beta),\\
y_3'&=&-\lambda_1 y_3+\tilde{f}_3(\zh{y},\,\beta),
\end{eqnarray}
where $\tilde{f}_i$, $i=1,\,2,\,3$, are such that $\partial
  \tilde{f}_i/\partial y_j=0$, $i,\,j=1,\,2,\,3$, at $\zh{y}=\zh{y}_f$. After such a change of variables, the origin is a stationary point corresponding to $\zh{y}_f$ and sufficiently close to the origin, the terms $\tilde{f}_1(\zh{y},\,\beta)$, $\tilde{f}_2(\zh{y},\,\beta)$ and $\tilde{f}_3(\zh{y},\,\beta)$ are negligibly small, so that near the origin the dynamical system can be approximated by the linearised system
\begin{eqnarray}
y_1'&=&\lambda_2 y_1-\omega y_2,\\
y_2'&=&\omega y_1+\lambda_2 y_2,\\
y_3'&=&-\lambda_1 y_3.
\end{eqnarray}

Let $\Sigma_1$ be a plane normal to the stable manifold of $\zh{y}_f$, $\Gamma_1$, and located at a small distance $\varepsilon_1$ from $\zh{y}_f$, {\it i.e.}, locally $\Sigma_1$ is given by 
\begin{equation}
\Sigma_1=\{(y_1,\,y_2,\,\varepsilon_1):\, y_1,\,y_2\in\mathbb{R}\}.
\end{equation}

Let $\Sigma_2$ be part of a plane transversal to the unstable manifold of $\zh{y}_f$, $\Gamma_2$, at some point near $\zh{y}_f$  and passing through $\zh{y}_f$ that is locally given by  
\begin{equation}
\Sigma_2=\{(y_1,\,0,\,y_3):\, |y_1-r^*|\leq \varepsilon_2,\, |y_3|\leq \varepsilon_3\}.
\end{equation}
Here $(r^*,\,0,\,0)\in\Gamma_1$ is sufficiently close to the origin and $\varepsilon_3<\varepsilon_1$. We denote the upper half-plane of $\Sigma_2$, when $y_3>0$, by $\Sigma_2^+$, {\it i.e.}, $\Sigma_2^+=\{\zh{y}\in\Sigma_2:\, y_3>0\}$ and let $\Sigma_2^-=\Sigma_2\backslash\Sigma_2^+$. We choose $\varepsilon_2$ to be sufficiently small so that each trajectory crosses $\Sigma_2$ only once. It can be shown that this condition is satisfied when $\varepsilon_2<\tanh(\lambda_2\pi/\omega)\:r^*$.

Using cylindrical polar coordinates $(r,\,\theta,\,z)$, such that $y_1=r\cos\theta$, $y_2=r\sin\theta$ and $y_3=z$, the linearised dynamical system near the origin is given by
\begin{eqnarray}
r'&=&\lambda_2 r,\\
\theta'&=&\omega,\\
z'&=&-\lambda_1 z.
\end{eqnarray}
The solution is given by 
\begin{eqnarray}
r&=&r_0\mathrm{e}^{\lambda_2 x},\\
\theta&=&\theta_0+\omega x,\\
z&=&z_0\mathrm{e}^{-\lambda_1 x}.
\end{eqnarray}

In the cylindrical polar coordinates, $\Sigma_1$ is given by $z=\varepsilon_1$ and $\Sigma_2$ is given by
\begin{equation}
\Sigma_2=\{(r,\,0,\,z):\, |r-r^*|\leq \varepsilon_2,\, |z| \leq \varepsilon_3\}.
\end{equation}

Let $\varphi_x$ be the flow map for the linearised dynamical system. Also, let $S$ be the set in $\Sigma_1$ given by
\begin{equation}
S=\{\zh{y}\in\Sigma_1:\,\exists\: x\text{ such that } \varphi_x(\zh{y})\in\Sigma_2\}.
\end{equation}
Then we can define the map
\begin{equation}
\varphi:S\rightarrow\Sigma_2:\, \zh{y}\mapsto \varphi_x(\zh{y})\text{ for some }x>0.
\end{equation}
It can easily be checked that the image of $\varphi$ is in fact $\Sigma_2^+$. Also, it can be easily seen that the set $S$ is the so-called Shilnikov snake, a set bounded by two spirals, $s_1$ and $s_2$, given by
\begin{equation}
r=(r^*\pm\varepsilon_2)\mathrm{e}^{-\lambda_2 x},\quad \theta=-\omega x,\quad z=\varepsilon_1,
\end{equation} 
respectively, where $x\in[(1/\lambda_1)\log(\varepsilon_1/\varepsilon_3),\,\infty)$, and the following segment of a straight line:
\begin{eqnarray}
&\displaystyle r\in\left[(r^*-\varepsilon_2)\left(\frac{\varepsilon_3}{\varepsilon_1}\right)^{\lambda_2/\lambda_1},\,(r^*+\varepsilon_2)\left(\frac{\varepsilon_3}{\varepsilon_1}\right)^{\lambda_2/\lambda_1}\right],\,\,\,&\\ 
&\displaystyle \theta=\frac{\omega}{\lambda_1}\log\left(\frac{\varepsilon_3}{\varepsilon_1}\right),\quad z=\varepsilon_1.&
\end{eqnarray} 
Let $l_p=\Sigma_1\cap W_u(\zh{y}_p)$ be the intersection of the
two-dimensional unstable manifold of $\zh{y}_p$ and the plane
$\Sigma_1$, which is locally a straight line given for $\beta=\beta_0$
by the equations $\theta=\theta_p$ and $z=\varepsilon_1$, where
$\theta_p$ is some constant. As $\theta_p\,\mathrm{mod}\,\pi$
determines the direction of the line, we can choose without out loss
of generality,
\begin{equation}
\theta_p\in(-\pi+(\omega/\lambda_1)\log(\varepsilon_3/\varepsilon_1),\,(\omega/\lambda_1)\log(\varepsilon_3/\varepsilon_1)].
\end{equation}
Next,
let $l_n$, $n=1,\,2,\,\ldots,$ be the intersections of the  line $l_p$ with set $S$ such that $|l_1|>|l_2|>\cdots,$ where $|l_n|$ denotes the length of the segment $l_n$, $n=1,\,2,\,\ldots,$ see fig.~\ref{fig:phase_space}. We can see that $l_n$ is given by 
\begin{eqnarray}
&&r\in[(r^*-\varepsilon_2)\exp(-\lambda_2(\pi (n-1)-\theta_p)/\omega),\,\nonumber\\
&&\qquad (r^*+\varepsilon_2)\exp(-\lambda_2(\pi (n-1)-\theta_p)/\omega)],\\ 
&&\theta=\theta_p-\pi (n-1)=\theta_p\,\mathrm{mod}\,\pi,\quad z=\varepsilon_1.
\end{eqnarray}
Then, we find that $\varphi(l_n)$ is a segment of a line in $\Sigma_2$ given by
\begin{eqnarray}
&&r\in[(r^*-\varepsilon_2),\,(r^*+\varepsilon_2)],\\ 
&&\theta=0,\\ 
&&z=\varepsilon_1\exp(-\lambda_1(\pi (n-1)-\theta_p)/\omega).
\end{eqnarray}
Let $l_b=\Sigma_2\cap W_s(\zh{y}_b)$ be the intersection of the two-dimensional stable manifold of $\zh{y}_b$ and the plane $\Sigma_2$. Locally it is a segment of a straight line, and since manifolds $W_u(\zh{y}_f)$ and $W_u(\zh{y}_b)$ intersect transversely, this segment of the line is given for $\beta=\beta_0$ by parametric equations
\begin{equation}
r=r^*+a s,\quad \theta=0,\quad z=s,
\end{equation}
where $a$ is some constant and $s$ is a parameter changing from $-\varepsilon_3$ to $\varepsilon_3$. Note that we can choose $\varepsilon_3$ to be smaller than $\varepsilon_2/|a|$ so that the line $l_b$ intersects all the lines $\varphi(l_n),$ $n=1,\,2,\,\ldots,$ and we denote such intersections points by $\zh{y}_{b,n},$ $n=1,\,2,\,\ldots.$ Let us denote the preimages of these points with respect to map $\varphi$ by $\zh{y}_{p,n},$ $n=1,\,2,\,\ldots.$ Note that $\zh{y}_{p,n}\in l_n,$ $n=1,\,2,\,\ldots.$ Next, since for each $n=1,\,2,\,\ldots,$ point $\zh{y}_{p,n}$ belongs to the unstable manifold of $\zh{y}_p$, there is an orbit $\Gamma_{p,n}$ connecting $\zh{y}_p$ and $\zh{y}_{p,n}$. Also, by definition of point $\zh{y}_{p,n}$, it is mapped by the flow map $\varphi_x$ to point $\zh{y}_{b,n}$ and the `transition time' from $\zh{y}_{p,n}$ to $\zh{y}_{b,n}$ is approximately equal to $x=t_\mathrm{tr}=(\pi(n-1)-\theta_p)/\omega$. Note that the difference in `transition times' from $\zh{y}_{p,n}$ to $\zh{y}_{b,n}$ and from $\zh{y}_{p,(n+1)}$ to $\zh{y}_{b,(n+1)}$ tends to $\pi/\omega$ as $n$ increases. We denote the orbit connecting $\zh{y}_{p,n}$ with $\zh{y}_{b,n}$ by $\Gamma_{f,n}$. Finally, since  $\zh{y}_{b,n}$ for each $n=1,\,2,\,\ldots,$ point $\zh{y}_{p,n}$ belongs to the stable manifold of $\zh{y}_b$, there is an orbit $\Gamma_{b,n}$ connecting $\zh{y}_{b,n}$ and $\zh{y}_{b}$. We conclude that there is an infinite but countable number of subsidiary heteroclinic orbits connecting points $\zh{y}_p$ and $\zh{y}_b$ that are given by $\Gamma_{s,n}=\Gamma_{p,n}\cup\Gamma_{f,n}\cup\Gamma_{b,n},$ $n=1,\,2,\,\ldots.$ Moreover, the difference in `transition times' for two successive orbits $\Gamma_{s,n}$ and $\Gamma_{s,(n+1)}$ taken to get from plane $\Sigma_1$ to plane $\Sigma_2$ tends to $\pi/\omega$ as $n\rightarrow\infty.$  Q.E.D.

\begin{figure}
\centering
\includegraphics[width=1\hsize]{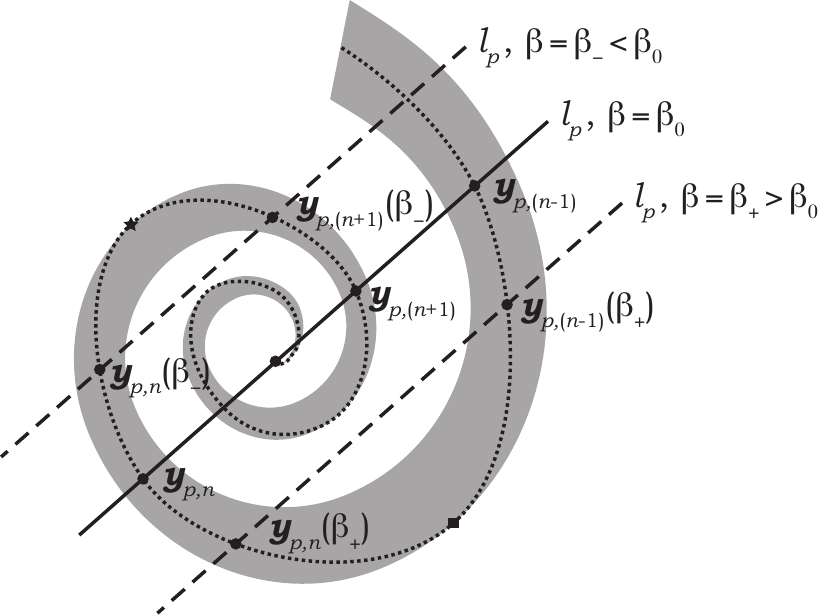}
\caption{Schematic representation of the Shilnikov snake, $S$, in plane $\Sigma_2$. The solid line shows line $l_p$ for $\beta=\beta_0$, the dashed lines show lines $l_p$ for $\beta=\beta_+>\beta_0$ and for $\beta=\beta_{-}<\beta_0$. The dotted line shows the locus of the points through which heteroclinic orbits connecting $\zh{y}_p$ and $\zh{y}_b$ pass for certain values of $\beta$ near $\beta_0$. The black square corresponds to the value of $\beta_+$ at which line $\l_p$ is tangent to $S$ and at which points $\zh{y}_{p,(n-1)}$ and $\zh{y}_{p,n}$ vanish in a saddle-node bifurcation. The star corresponds to the value of $\beta_-$ at which line $\l_p$ is tangent to $S$ and at which points $\zh{y}_{p,n}$ and $\zh{y}_{p,(n+1)}$ vanish in a saddle-node bifurcation.}
\label{fig:phase_space_snake}
\end{figure}

\begin{figure}
\centering
\includegraphics[width=0.7\hsize]{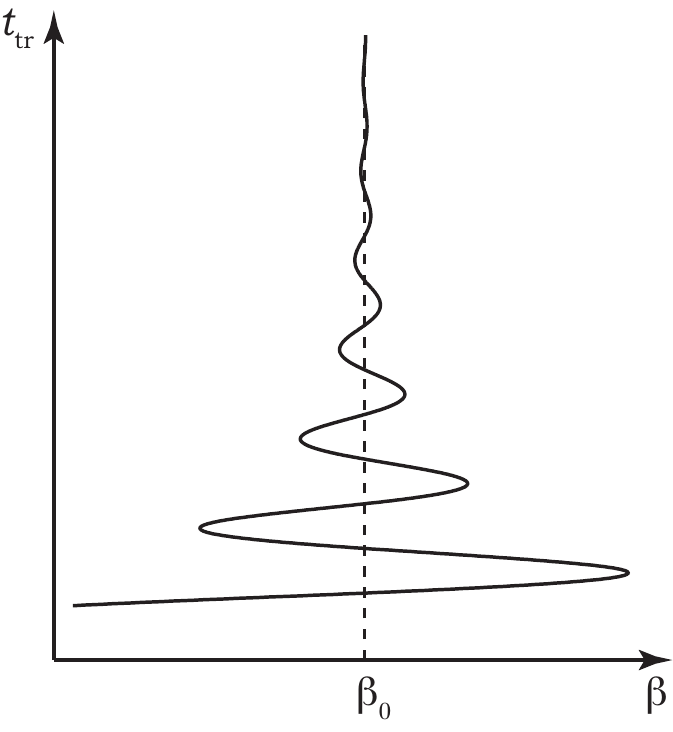}
\caption{Bifurcation diagram for heteroclinic orbits connecting $\zh{y}_p$ and $\zh{y}_b$.}
\label{fig:snaking}
\end{figure}

{\bf Remark.}  Snaking diagrams as those computed in the previous section are obtained by an unfolding of the structurally unstable heteroclinic chain connecting $\zh{y}_p$, $\zh{y}_f$ and $\zh{y}_b$. For $\beta$ close to $\beta_0$ but not necessarily equal to $\beta_0$, line $l_p=\Sigma_1\cap W_u(\zh{y}_p)$ is locally given by
\begin{equation}
y_2=b(\beta)y_1+c(\beta),\quad y_3=\varepsilon_1,
\end{equation}
where $c(\beta_0)=0$ and $b(\beta_0)=\tan(\theta_p)$ (without loss of generality, we can assume that $\theta_p\neq \pi/2+\pi n$ for any $n\in\mathbb{Z}$). This implies that in a small neighbourhood of point $(0,\,0,\,\varepsilon_1)$, this line can be approximated by
\begin{equation}
y_2=(b(\beta_0)+\Delta\beta\: b'(\beta_0))y_1+\Delta\beta\: c'(\beta_0),\quad y_3=\varepsilon_1,
\end{equation}
where $\Delta\beta=\beta-\beta_0$. Assuming that $c'(\beta_0)\neq 0$, we obtain that for $\beta\neq\beta_0$ line $l_p$ is shifted in plane $\Sigma_2$ and does not pass through point $(0,\,0,\,\varepsilon_1)$, see fig.~\ref{fig:phase_space_snake}. This implies that for $\beta\neq\beta_0$ line $l_p$ intersects the Shilnikov snake, $S$, finitely many times.  For sufficiently small  $\Delta\beta$, we denote by $\l_n(\beta)$ the intersection of $l_p$ with $S$ that is obtained by a small shift of $l_n$ for $\beta=\beta_0$. By considerations similar to those in the proof of the previous theorem, it can be shown that in each of the line segments there is a point $\zh{y}_{p,n}(\beta)$ such that there is a heteroclinic orbit passing through this point and connecting $\zh{y}_p$ and $\zh{y}_{b}$. For $\beta\neq\beta_0$ there is only a finite number of such orbits. Figure~\ref{fig:phase_space_snake} schematically shows $l_p$ by a solid line for $\beta=\beta_0$ and by dashed lines for $\beta=\beta_+>\beta_0$ and $\beta=\beta_-<\beta_0$. In addition, points $\zh{y}_{p,(n-1)}(\beta_+)$, $\zh{y}_{p,n}(\beta_+)$, $\zh{y}_{p,n}(\beta_-)$ and $\zh{y}_{p,(n+1)}(\beta_-)$ are shown. For certain value of $\beta_+$, points $\zh{y}_{p,(n-1)}(\beta_+)$, $\zh{y}_{p,n}(\beta_+)$ vanish in a saddle-node bifurcation. This point is indicated by a black square in the figure. At this point, line $l_p$ is tangent to the boundary of $S$. Also, for certain value of $\beta_-$, points $\zh{y}_{p,n}(\beta_-)$, $\zh{y}_{p,(n+)}(\beta_-)$ vanish in a saddle-node bifurcation. This point is indicated by a star in the figure. At this point, line $l_p$ is tangent to the boundary of $S$. The locus of the points  through which heteroclinic orbits connecting $\zh{y}_p$ and $\zh{y}_b$ pass for certain values of $\beta$ near $\beta_0$ is shown by a dotted line. It can be seen that this line is a spiral, $s$, that belongs to $S$, passes through points $\zh{y}_{p,n}$ and is tangent between transitions from $\zh{y}_{p,n}$ to  $\zh{y}_{p,(n+1)},$ $n=1,\,2,\ldots,$ to the boundary of $S$ given by spiral $s_1$. It can therefore be concluded that the bifurcation diagram showing the `transition time' for heteroclinic orbits connecting $\zh{y}_{p}$ and $\zh{y}_{p}$ versus parameter $\beta$ is a snaking curve, shown schematically in fig.~\ref{fig:snaking}, similar to the numerically obtained cases in figs.~\ref{fig:FIG2}, \ref{fig:FIG3} and \ref{fig:FIG4} for $\alpha=0.5$.  There is an infinite number of such orbits in a neighbourhood of $\beta_0$ and there is an infinite but countable number of saddle-node bifurcations that correspond to the points at which spiral $s$ touches the boundary of  the Shilnikov spiral, $S$.

We can find that the slope of the line tangent to spiral $s_1$ is
\begin{equation}
\frac{\mathrm{d}y_2}{\mathrm{d}y_1}=R \tan(\theta+\theta_0),
\end{equation}
where $R=\sqrt{\lambda_2^2+\omega^2}$ and $\theta_0=\tan^{-1}(\omega/\lambda_2)$. Therefore, at the points where line $l_p$ touches spiral $s_1$, we must have
\begin{equation}
R \tan(\theta_n+\theta_0)=b(\beta_0)+\Delta\beta_n b'(\beta_0),
\end{equation}
where $\theta_n$ and $\Delta\beta_n$ are the values of $\theta$ and $\Delta\beta$ corresponding to the $n^\text{th}$ saddle-node bifurcation. Thus, at these points
\begin{equation}
\theta_n=\tan^{-1}\left(\frac{b(\beta_0)}{R}+\Delta\beta_n\frac{b'(\beta_0)}{R}\right)-\theta_0-\pi n,
\end{equation}
for sufficiently large integer $n$. Equivalently,
\begin{equation}
x_n=-\frac{1}{\omega}\tan^{-1}\left(\frac{b(\beta_0)}{R}+\Delta\beta_n\frac{b'(\beta_0)}{R}\right)+\frac{\theta_0}{\omega}+\frac{\pi}{\omega} n.
\end{equation}
From this formula, we clearly see that the difference in tran\-si\-tion times between two saddle-node bifurcations tends to $\pi/\omega$. Also, at the saddle-node bifurcations we must have
\begin{equation}
\!r_n\sin\theta_n\!=\!(b(\beta_0)+\Delta\beta_n b'(\beta_0))r_n\cos\theta_n+\Delta\beta_n c'(\beta_0),\!\!\!\!
\end{equation}
where $r_n=(r^*+\varepsilon_2)\mathrm{e}^{-\lambda_2 x_n}$, which implies
\begin{equation}
\Delta\beta_n = r_n\frac{\sin\theta_n-b(\beta_0)\cos\theta_n} {c'(\beta_0)+b'(\beta_0)r_n}.
\end{equation}
From the latter expression, we can conclude that
\begin{equation}
|\Delta\beta_n|=O(r_n)=O(\mathrm{e}^{-\lambda_2 x_n}),
\end{equation}
which shows that the snaking bifurcation diagram approaches the vertical asymptote at an exponential rate, and explains the results presented in the bottom right panel of fig.~\ref{fig:FIG2} and in table~\ref{tab:3}.

Also, note that if $\zh{y}_f$ is a saddle, then the set $S$ is not a spiral but is a wedge-shaped domain. The line $l_p$ then passes through the vertex of this domain for $\beta=\beta_0$ and, generically, intersects it in the neighbourhood of the vertex only for $\beta<\beta_0$ but not for $\beta>\beta_0$ or vice versa. Then, the bifurcation diagram showing the `transition time' for heteroclinic orbits connecting $\zh{y}_{p}$ and $\zh{y}_{b}$ versus parameter $\beta$ is a monotonic curve instead of a snaking curve shown in fig.~\ref{fig:snaking}, similarly to the case in fig.~\ref{fig:FIG2} for $\alpha=0.1$.

{In the drawn meniscus problem the difference in
  tran\-si\-tion times between two saddle-\-node bifurcations (that
  tends to $\pi/\omega$) is a measure of the wavelength of the
  undulations on the foot and is therefore equivalent to the measures
  $\Lambda_{f}$ (as extracted from the steady thickness profiles) and
  $\Lambda_{s}$ (as extracted from the bifurcation curve) discussed in
  section~4 (see, in particular, table~5 where $\Lambda$ represents
  $\pi/\omega$). The overall transition time corresponds to the foot
  length $l_{f}$. Thus one can conclude that the bifurcation diagrams
  presented in figs.~\ref{fig:FIG2} and \ref{fig:FIG4} are explained
  by the results that have been presented in this section.}

\section{Conclusions}

We have analysed a liquid film that is deposited from a liquid bath
onto a flat moving plate that is inclined at a fixed angle to the
horizontal and is removed from the bath at a constant speed. {We
  have analysed a two-dimensional situation with a long-wave equation}
that is valid for small inclination angles of the plate and under the
assumption that the longitudinal length scale of variations in the
film thickness is much larger than the typical film thickness. {The
model equation used in most parts of our work includes} the terms due
to surface tension, the disjoining (or Derjaguin) pressure modelling
wettability, the hydrostatic pressure and the lateral driving force
due to gravity,
the dragging by the moving plate. {To further illustrate a
  particular finding we have also considered the situation where an
  additional lateral Marangoni shear stress results from a linear
  temperature gradient along the substrate direction.  Our main goal
  has been to analyse selected steady-state film thickness profiles
  that are related to collapsed heteroclinic snaking.}

{First, we have used centre manifold theory to properly derive
  the asymptotic boundary conditions on the side of the bath. In
  particular,} we have obtained asymptotic expansions of solutions in
the bath region, when $x\rightarrow\infty$. We found that in the
absence of the temperature gradient, the asymptotic expansion for the
film thickness, $h$, has the form $h\sim\sum_{n=-1}^\infty D_n
x^{-n}$, where without loss of generality $D_0$ can be chosen to be
zero (fixing the value of $D_0$ corresponds to breaking the
translational invariance of solutions and allows selecting a unique
solution from the infinite family of solutions that are obtained from
each other by a shift along the $x$-axis). In the presence of the
temperature gradient, this asymptotic expansion is not valid, but
instead consists of terms proportional to $x$, $\log x$ and
$x^{-m}\log^n x$, where $m$ and $n$ is a positive and a non-negative
integer, respectively. {Note that our systematically obtained
  sequence differs from the one employed in ref.~\cite{ME05}.}

Next, we have obtained numerical solutions of the ste\-a\-dy-\-state
equation and have analysed the behaviour of selected solutions as the plate
velocity and the temperature gradient are changed. When changing the
plate velocity, we observe that the bifurcation curves exhibit
collapsed heteroclinic snaking when the plate inclination angle is
larger than a certain critical value, namely, they oscillate around a
certain limiting velocity value, $U_\infty$, with an exponentially
decreasing oscillation amplitude and a period that tends to some
constant value. In contrast, when the plate inclination angle is
smaller than the critical value, the bifurcation curve is monotonic
and the velocity tends monotonically to $U_\infty$.
The solutions along these bifurcation curves are characterised by a
foot-like structure that emerges from the meniscus and is preceded by
a very thin precursor film further up the plate. The length of the
foot increases continuously as one follows the bifurcation curve as it
approaches $U_\infty$. {It is important to note that these
  solutions of diverging foot length do not converge to the
  Landau-Levich film solution at the same $U=U_\infty$. Indeed, the
  foot height at $U_\infty(\alpha)$ scales as $U^{1/2}$ while the
  Landau-Levich films scale as $U^{2/3}$. As expected, the results for
  the bifurcation curves that we here obtained with a precursor film
  model are similar to results obtained for such situations employing
  a slip model \cite{SADF07,ZiSnEg09}. The protruding foot structure
  has been observed in experiments, {\it e.g.}, in refs.~
  \cite{SADF07,DFSA08,Snoe08} where even an unstable part of the
  snaking curve was tracked. However, the particular transition
  described here has not yet been experimentally studied. This is in
  part due to the fact that in an experiment with a transversal
  extension (fully three-dimensional system) transversal meniscus and
  contact line instabilities set in before the foot length can
  diverge. We believe that experiments in transversally confined
  geometries may allow one to approach the transition more
  closely. Experiments with driving temperature gradients exist as
  well but focus on other aspects of the solution structure like, for
  instance, various types of advancing shocks (travelling fronts) and
  transversal instabilities \cite{BMFC1998prl}.  We are not aware of
  studies of static foot-like structures in systems with temperature
  gradients.}

{We further note that the described monotonic and non-monotonic
  divergence of foot length with increasing plate velocity may be seen
  as a dynamic equivalent of the equilibrium emptying transition
  described in ref.~\cite{PRJA2012prl}. There, a meniscus in a tilted
  slit capillary develops a tongue (or foot) along the lower wall. Its
  length diverges at a critical slit width. In our case, the length of
  the foot diverges at a critical plate speed -- monotonically below
  and oscillatory above a critical inclination angle. The former case
  may be seen as a continuous dynamic emptying transition with a close
  equilibrium equivalent. The latter may be seen as a discontinuous
  dynamic emptying transition that has no analogue at
  equilibrium. This is further analysed in ref.~\cite{GTLT14}.}

Finally, we have shown that in an appropriate three-dimensional phase
space, the three regions of the film profile, {\it i.e.}, the precursor film,
the foot and the bath, correspond to three fixed points, $\zh{y}_p$,
$\zh{y}_f$ and $\zh{y}_b$, respectively,  of a suitable dynamical
system. We have explained that the snaking behaviour of the
bifurcation curves is caused by the existence of a heteroclinic chain
that connects $\zh{y}_p$ with $\zh{y}_f$ and $\zh{y}_f$ with
$\zh{y}_b$ at certain parameter values. We have proved a general
result that implies that if the fixed points corresponding to the foot
and to the bath have two-dimensional unstable and two-dimensional
stable manifolds, respectively, and the fixed point corresponding to
the foot is a saddle-focus so that the Jacobian at this point has the
eigenvalues $-\lambda_1$, $\lambda_2\pm\mathrm{i}\:\omega$, where
$\lambda_{1,2}$ and $\omega$ are positive real numbers, then in the
neighbourhood of the heteroclinic chain there is an infinite but countable
number of heteroclinic orbits connecting the fixed point for the
precursor film with the fixed point for the bath. These heteroclinic
orbits correspond to solutions with feet of different
lengths. Moreover, these solutions can be ordered so that the
difference in the foot lengths  tends to $\pi/\omega$. We have also
explained that in this case the bifurcation curve shows a snaking
behaviour. Otherwise, if the fixed point corresponding to the foot is
a saddle, the Jacobian at this point has three real non-zero
eigenvalues, and the bifurcation curve is monotonic.

{The presented study is by no means exhaustive. It has focused
  on obtaining asymptotic expansions of the solutions in the bath
  region using rigorous centre manifold theory and on analysing the
  collapsed heteroclinic snaking behaviour associated with the dragged
  meniscus problems. However, the system has a much richer solution
  structure. Beside the studied solutions one may obtain Landau-Levich
  films and investigate their coexistence with the discussed foot and
  mensicus solutions. For other solutions the bath connects directly
  to a precursor-type film which then connects to a thicker
  `foot-like' film which then goes back to the precursor-type film
  that continues along the drawn plate. These solutions and their
  relation to the ones studied here will be presented elsewhere.}

\section*{Acknowledgements}
The authors acknowledge several interesting discussions about the
dragged film system with Edgar Knobloch, Serafim Kalliadasis, Andreas M\"unch, and Jacco
Snoeijer, and about emptying and other unbinding transitions with Andy
Parry and Andy Archer.  This work was supported by the European Union under grant
PITN-GA-2008-214919 (MULTIFLOW). The work of D.T. was partly 
supported by the EPSRC under grant EP/J001740/1. The authors are grateful to the
Newton Institute in Cambridge, UK, for its hospitality during a brief
common stay at the programme ``Mathematical Modelling and Analysis of
Complex Fluids and Active Media in Evolving Domains".

\bibliographystyle{unsrt}
\bibliography{bibliography}

\begin{thebibliography}{10}
\bibitem{noteManifold}
Here, the unstable manifold of $\boldsymbol{y}_p$ refers to the set of
  points $\boldsymbol{y}_0$ such that $\phi_t(\boldsymbol{y}_0)\rightarrow
  \boldsymbol{y}_p$ as $t\rightarrow-\infty$, where $\phi_t$ is the solution
  (or evolution) operator for the given dynamical system, and the stable
  manifold of $\boldsymbol{y}_b$ refers to the set of points $\boldsymbol{y}_0$
  such that $\phi_t(\boldsymbol{y}_0)\rightarrow \boldsymbol{y}_b$ as
  $t\rightarrow\infty$. These definitions are consistent with those given, {\it
  e.g.}, in ref.~\cite{Robi2004}.
\end{thebibliography}

\end{document}